\documentclass[apj]{emulateapj}

\usepackage{graphicx}% Include figure files

\usepackage{bm}% bold math
\newcommand{\qm}[1]{``#1''}

\def\sss{\scriptscriptstyle}
\def\U{{\sss \!U}}
\def\L{{\sss \!L}} 
\def\K{{\sss \!K}}

\def\P{{\sss \!P}}
\def\S{{\sss \!S}}

\def\R{{\sss \!R}}
\def\one{{\sss \!1}}
\def\two{{\sss \!2}}  
\def\nur{\nu_\mathrm{r}}
\def\nuv{\nu_\theta}
\def\nuL{\nu_\L}
\def\nuU{\nu_\U}
\def\nuK{\nu_\K}

\def\nul{\nuL}
\def\nuu{\nuU}

\def\nuP{\nu_\P}

\def\B{\mathcal{F}}

\begin{document}

%--------------------------------------------------------
%--------------------------------------------------------
 
\title{MASS--ANGULAR-MOMENTUM RELATIONS IMPLIED BY MODELS OF  TWIN PEAK QUASI-PERIODIC OSCILLATIONS}

\shorttitle{Mass--angular-momentum Relations Implied by NS kHz QPO Models}

%--------------------------------------------------------

\shortauthors{T\"or\"ok et al.}

\author{Gabriel T\"or\"ok, Pavel Bakala, Eva \v{S}r\'{a}mkov\'{a}, Zden\v{e}k Stuchl\'ik, Martin Urbanec,and Kate\v{r}ina Goluchov\'a}
 
\affil{Institute of Physics, Faculty of Philosophy and Science, Silesian University in Opava\\
               Bezru\v covo n\' am. 13, CZ-746\,01 Opava, Czech Republic\\www.physics.cz\\
               }
\email{Mailto: gabriel.torok{@}gmail.com}               

\date{\emph{Received 2010 November 30; accepted 2012 September 26; published 2012 November 16}}% It is always \today, today,
             
%--------------------------------------------------------

%--------------------------------------------------------
\begin{abstract}
{
Twin peak quasi-periodic oscillations (QPOs) appear in the X-ray power-density spectra of several accreting low-mass neutron star (NS) binaries. Observations of the peculiar Z-source Circinus X-1 display unusually low QPO frequencies. Using these observations, we have previously considered the relativistic precession (RP) twin peak QPO model to estimate the mass of central NS in Circinus X-1. We have shown that such an estimate results in a specific mass--angular-momentum ($M$--$j$) relation rather than a single preferred combination of $M$ and $j$. Here we confront our previous results with another binary, the atoll source 4U~1636--53 that displays the  twin peak QPOs at very high frequencies, and extend the consideration to various  twin peak QPO models.  In analogy to the RP model, we find that these imply their own specific $M$--$j$ relations. We explore these relations for both sources and note differences in the $\chi^2$ behavior that represent a dichotomy  between high- and low-frequency sources. Based on the RP model, we demonstrate that this dichotomy is related to a strong variability of the model predictive power across the frequency plane. This variability naturally comes from  the radial dependence of characteristic frequencies of orbital motion.
As a consequence, the restrictions on the models resulting from observations of  low-frequency sources are weaker than those in the case of high-frequency sources. Finally we also discuss the need for a correction to the RP model and consider the removing of $M$--$j$ degeneracies, based on the twin peak QPO-independent angular momentum estimates.}
\end{abstract}

%--------------------------------------------------------   


%--------------------------------------------------------

.%---------------------------------------------------------
\section{Introduction}
%---------------------------------------------------------
\label{section:introduction}

Several low-mass neutron star binaries (NS LMXBs) exhibit in the high-frequency part of their X-ray power-density spectra (PDS) two distinct peaks, so-called  twin peak quasi-periodic oscillations (QPOs). The two peaks are referred to as the upper and lower QPO.   {{Centroid frequencies of these QPOs, $\nuL$ and $\nuU$, vary over time, but follow  frequency correlations specific to individual sources. However, these specific correlations are qualitatively similar \citep[see][and references therein]{psa-etal:1999a,ste-etal:1999,abr-etal:2005a,abr-etal:2005b}.} In some cases, the frequency ranges spanned by a single source are as large as a few hundreds of Hz. At present, there is no consensus on the QPO origin. Numerous models have been proposed, mostly assuming that the two twin QPOs carry important information about the inner accreting region dominated by the effects of strong Einstein's gravity. In principle, several of these models imply restrictions to netron star (NS) parameters \citep[a systematic treatment of these restrictions through the fitting of twin peak QPO correlations was pioneered by][]{psa-etal:1998}. A brief introduction to QPOs and their models can be found in \cite{Kli:2006:CompStelX-Ray:}.} 

In our previous work, \cite{tor-etal:2010}, hereafter Paper I, we focused on restrictions of a particular \qm{relativistic precession} (RP) QPO model and a peculiar bright Z-source Circinus X-1.  {The RP model introduced by \citet{ste-vie:1999} and \citet{ste-etal:1999}} identifies the lower and upper kHz QPOs with the periastron precession $\nuP$ and Keplerian $\nuK$ frequency of a perturbed circular geodesic motion at the given radii $r$,
%---------------------------------------------------------
\begin{equation}
\label{equation:stella}
\nul(r)=\nuP(r)=\nuK(r)-\nur(r),\quad\nuu(r)=\nuK(r),
\end{equation}
%---------------------------------------------------------
where $\nur$ is the radial epicyclic frequency of the Keplerian motion. In Paper~I we noticed that the RP model well matches the data points of Circinus X-1 for any dimensionless NS angular momentum, $j\equiv cJ/GM^2$, when the assumed NS mass reads $M\sim 2.2M_\sun[1+0.55(j+j^2)]$. We have shown that the existence of such a mass--angular-momentum $(M-j)$ relation is generic for the model.

Circinus X-1 that we discussed in Paper~I is a relatively well-known source, since it displays twin QPOs at unusually low frequencies, $\nuL\in(50{\mathrm{Hz}},~250{\mathrm{Hz}})$ and $\nuU\in(200{\mathrm{Hz}},~500{\mathrm{Hz}})$ \citep[see][who discovered its QPOs]{bou-etal:2006}. Here we consider another binary, a faint atoll source 4U~1636--53 that, on the contrary, displays  twin peak QPOs at very high frequencies, $\nuL\in(550{\mathrm{Hz}},~1000{\mathrm{Hz}})$ and $\nuU\in(800{\mathrm{Hz}},~1250{\mathrm{Hz}})$ \citep[][]{bar-etal:2005a,bar-etal:2005b}. As illustrated in Figure~\ref{figure:stella}(a), also assuming this source we can confront the two representatives of {high-} and {low-}frequency  twin peak QPO sources (in general,  twin peak QPOs are more often detected at rather low frequencies in Z-sources and high frequencies in atoll sources, but there are some counterexamples, e.g., Z-source Sco~X-1; see \cite{Kli:2006:CompStelX-Ray:}). Apart from the RP model, we extend our consideration to several other  twin peak QPO models. The text is organized as follows.

In Section~\ref{section:introduction:Mj}, we briefly recall some points from the Paper~I that are of generic importance for the present work. In Section~\ref{section:models}, we briefly recall the data used and their origin along with the set of QPO models that are considered within the paper. %We also shortly comment on the frequency relations $\nuU(\nuL)$ predicted by individual models.
In Section~\ref{section:results}, we fit the data points with the frequency relations predicted by individual models and show that, in analogy to the RP~model, each of them implies its specific mass--angular-momentum relation. In Section~\ref{section:dichotomy}, we discuss {the issue of the models predictive power variability across the frequency plane. We also briefly investigate} the requirement of a correction to the RP model and suggest that it can be relevant to both high- and low-frequency sources. In Section~\ref{section:conclusions}, we discuss our results and present some concluding remarks.

%--------------------------------------------------------
\section{Mass--Angular-Momentum Relation from RP Model}
%--------------------------------------------------------
\label{section:introduction:Mj}

As found in Paper~I, the data points of Circinus X-1 are well matched by the RP model when the combinations of the source mass and angular momentum are in the form of $M\sim 2.2M_\sun[1+0.55(j+j^2)]$. The existence of such an $(M-j)$ relation is generic to the model. Let us briefly recall the major points and implications of our previous results.

We found that due to the properties of the RP model and the NS spacetime the quality of  fit for a given source should not differ much from the following general relation:
%--------------------------------------------------------
\begin{equation}
\label{equation:general}
M\sim M_0\left[1+k\left(j+j^2\right)\right]\,.
\end{equation}
%--------------------------------------------------------
In this relation, $M_0$ is the mass that provides the best fit assuming a non-rotating star $(j=0)$. The coefficient $k$, implied by the model, would read $k=0.7$ if the measured data points were sampled uniformly along the large range of frequencies
\begin{equation}
\nuL\in(\nu\ll \nu_\mathrm{ISCO},\,\nu_\mathrm{ISCO}),
\end{equation}
 where $\nu_\mathrm{ISCO}$ denotes the Keplerian orbital frequency at the innermost stable circular orbit $r_\mathrm{ms}$ (hereafter ISCO). The available data points are, however, unequally sampled and often cluster, either simply due to incomplete sampling and weakness of the two QPOs outside the limited frequency range, or due to the intrinsic source clustering \citep[][]{abr-etal:2003a, bel-etal:2005,bel-etal:2007b,  tor-etal:2008b, tor-etal:2008c, tor-etal:2008a, bar-bou:2008,  bou-etal:2010}.

%--------------------------------------------------------
\subsection{Importance of Frequency Ratio for the RP Model Predictions in Different Sources}
%--------------------------------------------------------
\label{section:introduction:ratio}

In the detailed analysis presented in  Appendix A.2 of Paper~I, we elaborated the influence of unequal sampling of the frequency correlation $\nuU(\nuL)$. It is important that frequencies predicted from the RP model scale as $1/M$ for a fixed $j$, and in this sense the expected frequency ratio, $R\equiv\nuU/\nuL$, is mass independent. Moreover, in the RP model, it is
%------------------------------------
\begin{equation}
\label{equation:ratio}
R=\nuK/\left(\nuK-\nur\right)\,.
%R=\frac{\nuK}{\nuK-\nur}\,.
\end{equation}
%-----------------------------
The frequency $\nur$ vanishes when the radial coordinate approaches ISCO, $r\rightarrow r_\mathrm{ms}$, and therefore $R\rightarrow1$. On the other hand, when $r\rightarrow\infty$ the spacetime becomes flat reaching Newtonian limit where $\nur\rightarrow\nuK$ and $R$ diverges. The QPOs that are expected to arise close to ISCO therefore always reveal a low $R$, while those expected to arise in a large radial distance from the NS reveal a high $R$. For any NS parameters, the top part (relatively high frequencies) of a given frequency correlation $\nuU(\nuL)$ predicted by the RP model then reveals a frequency ratio close to $R=1$. The bottom part (relatively low frequencies) of the frequency correlation reveals high frequency ratio $R\gtrsim3$.\footnote{In Paper~I we have shown that more than $60\%$ of the length of the expected curve $\nuU(\nuL)$ correspond to $R<3$ (see Figure~9 of Paper~I).}

Based on the above-mentioned theoretical prediction of the RP model, in Paper~I we found that the value of $k$ in mass--angular-momentum relation (\ref{equation:general}) must tend to $k\sim0.75$ when the range of the ratio of the lower and upper QPO frequencies  in the sample falls to low values close to $R=1$. On the other hand, it is $k\sim0.5$ when the range of $R$ has a high value ($R\sim5$). This consequence of unequal sampling does not depend on the absolute value of QPO frequencies. 

 {As noticed first by \citet{ste-vie:1999} and \citet{ste-etal:1999} and later discussed in several works \citep[e.g.,][]{bel-etal:2007a}}, frequencies of the twin peak QPOs observed in most of the NS sources are roughly matched by the frequency correlation implied by the RP model for the NS mass $M\sim2M_\sun$. Assuming this mass, the low frequency ratio $R\lesssim1.5$ roughly corresponds to high QPO frequencies, $\nul\sim0.6-1$kHz, while $R\sim(2-5)$ corresponds to low QPO frequencies, $\nul\sim50-500$Hz. This roughly matches the phenomenological division between \qm{low-} and \qm{high}-frequency  twin peak QPO sources based on distribution of typical frequencies of QPOs observed in individual continuous observations. Thus, in practice, the expected value of $k=0.7$ changes due to unequal sampling only very slightly to $k\sim0.7-0.75$ for available data on {high-frequency}  twin peak QPO sources. For the available data on {low-frequency}  twin peak QPO sources, the effect of unequal sampling is more important, changing $k$ to $\sim0.5-0.65$, which also corresponds  to the case of Circinus X-1 elaborated in Paper~I. Detailed quantification of restrictions on $k$ can be found in Table~1 of Paper~I.

Next we justify our result comparing the case of Circinus X-1 to the case of high-frequency source 4U~1636$-$53, for which we expect $k\sim0.7-0.75$. Then we explore whether several other QPO models imply their own $M-j$ relations or not. 

%--------------------------------------------------------------
\section{Data and Models}
%--------------------------------------------------------
\label{section:models}

Figure~\ref{figure:stella}(a) shows several  twin peak QPO data points coming from the works of \citet{bar-etal:2005a,bar-etal:2005b, Boirin-00,diSalvo-03,Homan-02,Jonker2002a,Jonker2002b,MvdK2000,M2001,van-Straaten-00,van-Straaten-02,Zhang-98}, and \cite{bou-etal:2006}. For the analysis presented in this paper we use the  twin peak QPO data of 4U~1636-53 \citep[from][]{bar-etal:2005a,bar-etal:2005b} and Circinus X-1 \citep[from][]{bou-etal:2006}. These data points are denoted in the figure by the color-coded symbols. Each of them corresponds to an individual continuous segment of the source observation.  {One can see that our choice of the two representative NSs allows us to demonstrate the confrontation between the low- and high-frequency sources, as mentioned in the previous section.} Details of the observations, data analysis techniques, and properties of the twin peak QPOs in the two sources discussed can be found in \cite{bar-etal:2005a,bar-etal:2005b,bar-etal:2006,bou-etal:2006,men:2006}, and \cite{Kli:2006:CompStelX-Ray:}. 

Each of the many QPO models proposed \citep[e.g.,][]{alp-sha:1985,lam-etal:1985,mil-etal:1998a,psa-etal:1999b,wag:1999,wag-etal:2001,abr-klu:2001,tit-ken:2002,rez-etal:2003,pet:2005a,zha:2005,kat:2007,stu-etal:2008,muk:2009} still faces several difficulties and, at present, none of them is favored. In such a situation, we expect that the estimations of mass and angular momentum based on the individual models could be helpful for the further development or falsification of an appropriate model. In the next section we therefore consider several of these models in addition to the RP model investigated in Paper~I, and examine what mass--angular-momentum relations they imply. Since we do not attempt to describe  the individual models and resolve all their specific issues in detail, in what follows we just give  a short summary of the models examined and highlight some of their distinctions along with the related references.\footnote{Some more details on these models and a discussion of their relevance to the black-hole QPOs can be found in \cite{tor-etal:2011}.}

%--------------------------------------------------------------
\subsection{Individual Models}
%--------------------------------------------------------------

The RP model  has been proposed in a series of papers by \cite{ste-vie:1998a,ste-vie:1998b,ste-vie:1999,ste-vie:2002} and \cite{mor-ste:1999} and explains the kHz QPOs as a direct manifestation of modes of relativistic epicyclic motion of blobs at various radii $r$ in the inner parts of the accretion disk. Within the model, the  twin peak QPO frequency correlation arises due to periastron precession of the relativistic orbits. Because of the existence of another so-called Lense--Thirring RP the model also predicts another frequency correlation extending to higher timescales.  {The kHz QPO frequencies are indeed correlated with the low-frequency QPO features observed far below 100 Hz, which was first noticed and discussed in the works of  \cite{psa-etal:1999a}, \cite{ste-vie:1999}, and \cite{ste-etal:1999}. Here we restrict our attention mostly to kHz features but the low-frequency QPO interpretation within the RP model is briefly considered in Section \ref{section:conclusions} and Appendix~\ref{section:appendix:RP:low}.}

Recently, \cite{cad-etal:2008}, \cite{kos-etal:2009}, and \cite{ger-etal:2009} have introduced a similar concept in which the QPOs were generated by a \qm{tidal disruption} (TD) of large accreting inhomogeneities. It is assumed--and is supported by some hydrodynamic simulations--that blobs orbiting the central compact object are stretched by tidal forces forming a \qm{ring-section} features that are responsible for the observed modulation. The model has been proposed for black hole (BH) sources (both supermassive and stellar mass) but, in principle, it should work for compact NS sources as well. In some cases at least, the PDS produced within the model seem to well reproduce those observed.

It is often argued that QPOs arise due to \qm{disk oscillations} (in contrast to the above models considering \qm{hot-spot motion}) and that some resonances can be involved. The disk-oscillation concept has a good potential for explaining the high QPO coherence times observed in some NS systems \citep[see][who first recognized the importance of the high QPO quality factor measured in 4U~1636--53, $Q\sim200$]{bar-etal:2005c}.  The resonance hypothesis is supported by the appearance of the 3:2 frequency ratio observed in BH sources \citep[][]{abr-klu:2001,mcc-rem:2006,tor-etal:2005}. There is also a less straightforward evidence for the importance of the same 3:2 ratio in the case of  NS sources which was first noticed in terms of the frequency ratio $R\equiv\nuU/\nuL$ clustering  \citep[see][for details and related discussion]{abr-etal:2003a, bel-etal:2005,bel-etal:2007b, tor-etal:2008b, tor-etal:2008c, tor-etal:2008a,  bou-etal:2010}. As found recently, in the six atoll NS systems including 4U~1636$-$53, the difference between the rms amplitudes of the upper and lower QPOs changes its sign for resonant frequency ratios $R=3\!:\!2$ \citep{tor:2009}. This interesting effect still requires some further  investigation, since the rms amplitudes of kHz QPOs are energy dependent and this must be taken into account. Nevertheless, we note that it was suggested by \citet{hor-etal:2009} that the \qm{energy switch} effect could be naturally explained in terms of the theory of the nonlinear resonance. 

Two examples of the often quoted resonant disk-oscillation models are the epicyclic resonance (ER) model \cite[][]{klu-abr:2001,abr-etal:2003b,abr-etal:2003c,klu-etal:2004} assuming axisymmetric modes and the \qm{warped disk} (WD) oscillation model suggested by \cite{kat:2001,kat:2007,kat:2008} that assumes non-axisymmetric modes. We consider these and also another two QPO resonance models dealing with different combinations of non-axisymmetric disk-oscillation modes. The latter two models are of particular interest because they involve oscillation modes whose frequencies almost coincide with the frequencies predicted by the RP model when the NS rotates slowly. We denote them as RP1 \citep{bur:2005} and RP2 \citep{tor-etal:2007,tor-etal:2010} models and assume that the resonant corrections to the eigenfrequencies are negligible.
%--------------------------------------------------------------
\subsubsection{Frequency Relations}
%--------------------------------------------------------------

The relations that define the upper and lower QPO frequencies in terms of the orbital frequencies are given for each of the above models in the first column of  Table~\ref{table:1}. {We include these terms for the case of the Kerr spacetimes in  Appendix~\ref{section:appendix:relations}. The applicability of an approach assuming the Kerr spacetimes for high-mass NSs was elaborated in Paper~I. The relevance and limitations of the same approach within the work and results presented here are discussed more in Section~\ref{section:conclusions} and Appendix~\ref{section:appendix:HT}.}

For the RP model, one can easily solve the {definition} relations to arrive at the explicit formula which relates the upper and lower QPO frequencies. A similar simple evaluation of an explicit relation between the two observed QPO frequencies is also possible  for the TD model. For the RP and TD models, we give the explicit formulas in Equations (\ref{equation:A:RP}) and (\ref{equation:A:TD}). For the WD, RP1, and RP2 models the definition relations lead to high-order polynomial equations that relate the lower and upper QPO frequencies. In these cases, in Appendix~\ref{section:appendix:relations} we give only the implicit form of the $\nuU(\nuL)$ function  which has to be treated numerically.

For the version of ER model assumed here, we expect  the $\nuU(\nuL)$ function  in the form of a linear relation. This approach follows the work of \cite{abr-etal:2005a,abr-etal:2005b} and related details are briefly recalled in Section~\ref{section:ER:model}.

%---------------------------------------------------------------------------------------------------
\section{Data Matching}
%---------------------------------------------------------------------------------------------------
\label{section:results}

In this section we fit the data points of 4U 1636--53 and Circinus X-1 with frequency relations predicted from each of the individual models (i.e., by functions (\ref{equation:A:RP}) and (\ref{equation:A:TD}) for the RP and TD model, respectively, by a straight line for the ER model, and by the numerically given  solutions of Equations~(\ref{equation:A:WD})--(\ref{equation:A:RP2})  for the other models).\footnote{At this point we should also note that our choice of models represents a subset of those recently discussed by \cite{lin-etal:2010} for the two sources 4U~1636--53 and Sco~X-1. An overlap with their work is discussed in Section~\ref{section:conclusions}.} As in Paper~I, we restrict the range of  mass and angular momentum considered to $[M\in(1,\,4)M_\sun]\times[j\in(0,\,0.5)]$. For all the models {except the ER model (Section~\ref{section:ER:model})}, we first find the best fit in the Schwarzschild spacetime ($j=0$) for a single free parameter $M$ using the least-squares fitting procedure \citep[e.g.,][]{pre-etal:2007}. Then we also inspect the two-dimensional $\chi^2$ behavior for the free $M$ and $j$.

{Within the numerical approach adopted the model frequency curve is parameterized along its full length through a parameter $p$ which ranges from $p_\infty$ to $p_\mathrm{ISCO}$. The exact definition of $\chi^2$ that we use here is then given as}
\begin{equation}
\label{equation:definition:chisq}
\chi^2\equiv\sum^{m}_{{n}=1} \Delta_{n}^2,~\mathrm{with}~\Delta_{n}=\mathrm{Min}\left(\frac{l_{{n},\,{p}}}{\sigma_{{n}\,,{p}}}\right)_{{p}_\infty}^{{p}_\mathrm{ISCO}},
\end{equation}
{where $l_{{n},\,{p}}$ is the length of a line between the $n$th measured data point $[\nul({n}),~\nuu({n})]$ and a point $[\nul({p}),~\nuu({p})]$ belonging to the model frequency curve. The quantity $\sigma_{{n}\,,{p}}$ equals the length of the part of this line located within the error ellipse around the data point.}  
%--------------------------------------------------------------
\subsection{Results for the RP, RP1, and RP2 Models}
%--------------------------------------------------------------

{ {Considering $j=0$ for fitting the data of 4U~1636--53 with the RP model, we find a narrow $\chi^2$ minimum for $M_0\sim1.8M_\sun$ but its value is rather high, $\chi^2\doteq {350/21} \mathrm{dof}$.} We also find that there is no sufficient improvement along the whole  given range of mass even up to the upper limit of $j$. Thus, assuming that the model is valid, we can only speculate that there is an unknown systematic uncertainty. { {Then it follows from  Equation~(\ref{equation:definition:chisq}) that the $\chi^2$ of the best fit for $j=0$ drops to an acceptable value $\chi^2=1 \mathrm{dof}$ when the uncertainties in the measured QPO frequencies are multiplied (underestimated) by factor $\xi\equiv\sqrt{\chi^2/\mathrm{dof}}\doteq4$. Under this consideration we find the NS mass from the best-fit reading $M_0=1.78M_\sun$. We express the corresponding scatter in the estimated mass as $\delta M=[\pm {0.03}]M_\sun$, assuming the $2\sigma$ confidence level which we henceforth use as the reference one.}

 {On the other hand, the best match to the data of Circinus~X-1 for the RP model and $j=0$ already reveals an acceptable value of $\chi^2\doteq {12.9/10} \mathrm{dof}$, and in summary, we can write the quantities $M_0$ inferred from the RP model for both sources as}
%--------------------------------------------------------------
\begin{eqnarray}
\label{equation:RP:fit:1636}
&&M_0=1.78[\pm {0.03}]M_\sun\nonumber\\
&&\mathrm{in~4U~1636-53}\quad(\chi^2=1 \mathrm{dof}\Leftrightarrow\xi\doteq4)\,\\
%\mathrm{and~(Paper~I)}&&\nonumber\\
\mathrm{and}&&\nonumber\\
\label{equation:RP:fit:circinus}
&&M_0= {2.19}[\pm {0.3}]M_\sun\nonumber\\
&&\mathrm{in~Circinus~X-1}\quad(\chi^2= {12.9/10} \mathrm{dof}). 
\end{eqnarray}
%--------------------------------------------------------------

As found in Paper~I and briefly recalled here in Section~\ref{section:introduction:Mj}, for the RP model and a given source the $\chi^2$ should not  differ much along the $M-j$ relation $M\sim M_0[1+k(j+j^2)]$ where $k\sim0.7-0.75$ for {high-}frequency sources and $k\sim0.5-0.6$ for the {low-}frequency sources.  The results of the two-dimensional fitting of the parameters $M$ and $j$ agree well with this finding. The $\chi^2$ behavior for 4U~1636--53 is depicted and compared to the case of Circinus X-1 in the form of color-coded maps in Figure~\ref{figure:stella}(b).  {Clearly, the best fits are reached when $M$ and $j$ are related through the specific relations denoted by the  dashed green lines. We approximate these relations in the form $M=M_0\times[1+k(j+j^2)]$ arriving at the following terms:}
%--------------------------------------------------------------
\begin{eqnarray}
\label{equation:RP:Mjfit:1636}
&&M=1.78[\pm {0.03}]M_\sun\times[1+ {0.73}(j+j^2)]\nonumber\\
&&\mathrm{in~4U~1636-53}\\
\mathrm{and}&&\nonumber\\
\label{equation:RP:Mjfit:circinus}
&&M= {2.19}[\pm {0.3}]M_\sun\times[1+ {0.52}(j+j^2)]\nonumber\\
&&\mathrm{in~Circinus~X-1}.
\end{eqnarray}
%--------------------------------------------------------------

%--------------------------------------------------------
%--------------------------------------------------------
%                           FIGURE
%--------------------------------------------------------
%--------------------------------------------------------
%---------------------------------------------------------
\begin{figure*}[t!]
(a)\hfill (b) \hfill~
\smallskip

\begin{minipage}{1\hsize}
\begin{center}
\includegraphics[width=.92\textwidth]{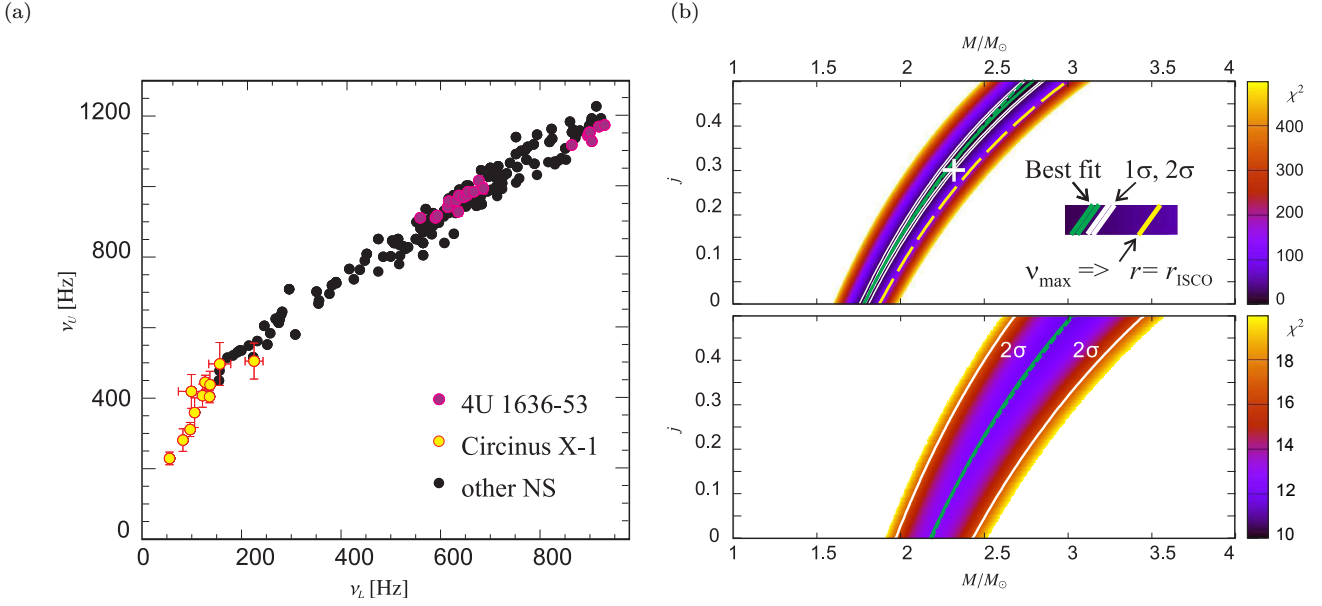}
\end{center}
\end{minipage}
\caption{ (a)  Twin peak QPO frequencies in the atoll source 4U 1636--53 (22 data points in purple), Z-source Circinus X-1 (11 data points in red/yellow), and several other atoll- and Z-sources (data points in black). (b) The $\chi^2$ dependence on  $M$ and $j$ for the RP model. The top panel corresponds to 4U~1636--53 while the bottom panel corresponds to Circinus X-1.  {For 4U~1636--53 $\xi=4$ is assumed}. The dashed green line indicates the best $\chi^2$ for a fixed $M$. The continuous green line denotes its quadratic approximation.  {The white lines indicate corresponding 1$\sigma$ and 2$\sigma$ confidence levels.} The white cross-marker denotes the mass and angular momentum reported for 4U~1636--53 and the RP model by (\cite{lin-etal:2010}; see Section~\ref{section:conclusions}).  {The dashed yellow line in the top panel indicates a simplified estimate on the upper limits on $M$ and $j$ assuming that the highest observed upper QPO frequency in 4U~1636--53 is associated with the ISCO. This estimate is not included for Circinus~X-1 because the observed frequencies clearly points to the radii far away from ISCO which can be seen from Figure \ref{figure:fits}.}
}  
\label{figure:stella}
\end{figure*}
%----------------------------------------%--------------------------------------------------------------
%----------------------------------------%--------------------------------------------------------------

%----------------------------------------%--------------------------------------------------------------
\subsubsection{Results for the RP1 Model}
%--------------------------------------------------------------

The frequencies predicted by the RP and RP1 models are very similar for slowly rotating NSs.  The two models commonly define the lower observable QPO frequency as 
%--------------------------------------------------------------
\begin{equation}
\label{equation:periastron}
\nuL=\nuK-\nur.
\end{equation}
%--------------------------------------------------------------
The upper observable QPO frequencies differ, reading 
%--------------------------------------------------------------
\begin{equation}
\nuU^\mathrm{~\R\P}=\nuK\,,\quad \nuU^\mathrm{~\R\P\one}=\nuv\,.
\end{equation}
%--------------------------------------------------------------
In the Schwarzschild limit $j=0$, $\nuv=\nuK$ and $\nuU$ is common to both RP and RP1. Consequently,
%--------------------------------------------------------------
\begin{eqnarray}
\label{equation:RP12:fits}
M_0^\mathrm{~\R\P\one}=M_0^\mathrm{~\R\P}\,,
\end{eqnarray}
%--------------------------------------------------------------
where $M_0^\mathrm{~\R\P}$ is given in Equation (\ref{equation:RP:fit:1636}) and (\ref{equation:RP:fit:circinus}) for 4U~1636--53 and Circinus X-1, respectively. {For 4U~1636-53, the quality of the fits does not differ much between $j=0$ and $j\neq0$ and the same conclusions on the possible unknown systematic uncertainty as in the case of {RP} model are valid.}

One can expect that fits to the data based on the RP1 model for $j\neq0$ should exhibit $M-j$ degeneracy qualitatively similar to the case of the RP model. We do not repeat for RP1 model the full analysis of $M-j$ degeneracy presented in the Paper~I for the RP model. Instead, we just inspect the behavior of $\chi^2$ for free $M$ and $j$ to check whether such degeneracy is present and evaluate it. The $\chi^2$ behavior resulting for free $M$ and $j$ is depicted in the form of color-coded maps in Figure~\ref{figure:maps}(a). The two  $\chi^2$ maps displayed clearly reveal $M-j$ degeneracy qualitatively similar to that of the RP model. Related $M-j$ relations (best $\chi^2$ for a fixed $M$) are denoted by dashed green lines in  Figure~\ref{figure:maps}(a).  {We approximate these relations in the form $M=M_0\times[1+k(j+j^2)]$ arriving at the following terms:}
%--------------------------------------------------------------
\begin{eqnarray}
\label{equation:RP1:Mjfit:1636}
&&M=1.78[\pm {0.03}]M_\sun\times[1+ {0.48}(j+j^2)]\nonumber\\
&&\mathrm{in~4U~1636-53}\\
\mathrm{and}&&\nonumber\\
\label{equation:RP1:Mjfit:circinus}
&&M= {2.19}[\pm {0.3}]M_\sun\times[1+ {0.39}(j+j^2)]\nonumber\\
&&\mathrm{in~Circinus~X-1}.
\end{eqnarray}
%--------------------------------------------------------------

%--------------------------------------------------------
%--------------------------------------------------------
%                           FIGURE
%--------------------------------------------------------
%---------------------------------------------------------
\begin{figure*}
(a)~~{\small{RP1 model}}\hfill ~~~~(b)~~{\small{RP2 model}}\hfill~
\smallskip

\begin{minipage}{1\hsize}
\begin{center}
\includegraphics[width=.92\textwidth]{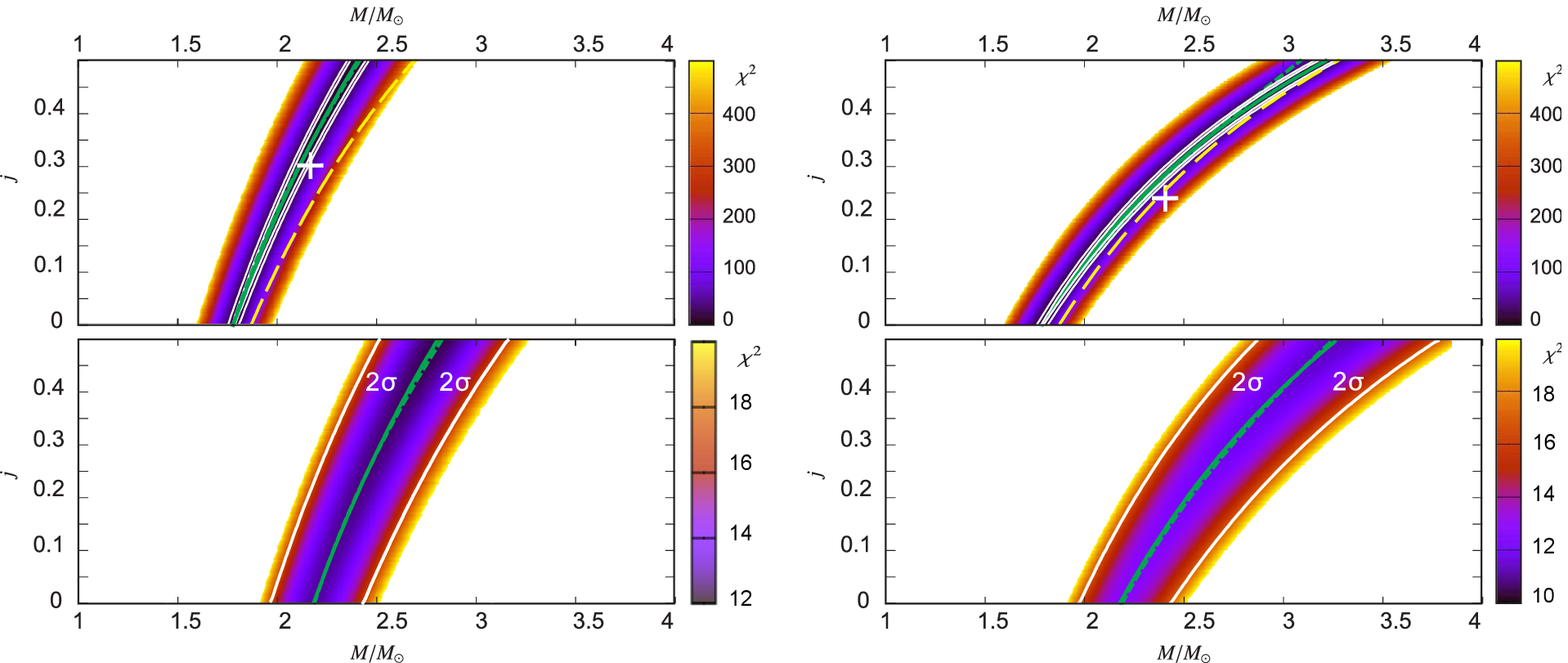}
\end{center}
\end{minipage}
\medskip

(c)~~{\small{WD model}}\hfill ~~~~(d)~~{\small{TD model}}\hfill~
\smallskip

\begin{minipage}{1\hsize}
\begin{center}
\includegraphics[width=.92\textwidth]{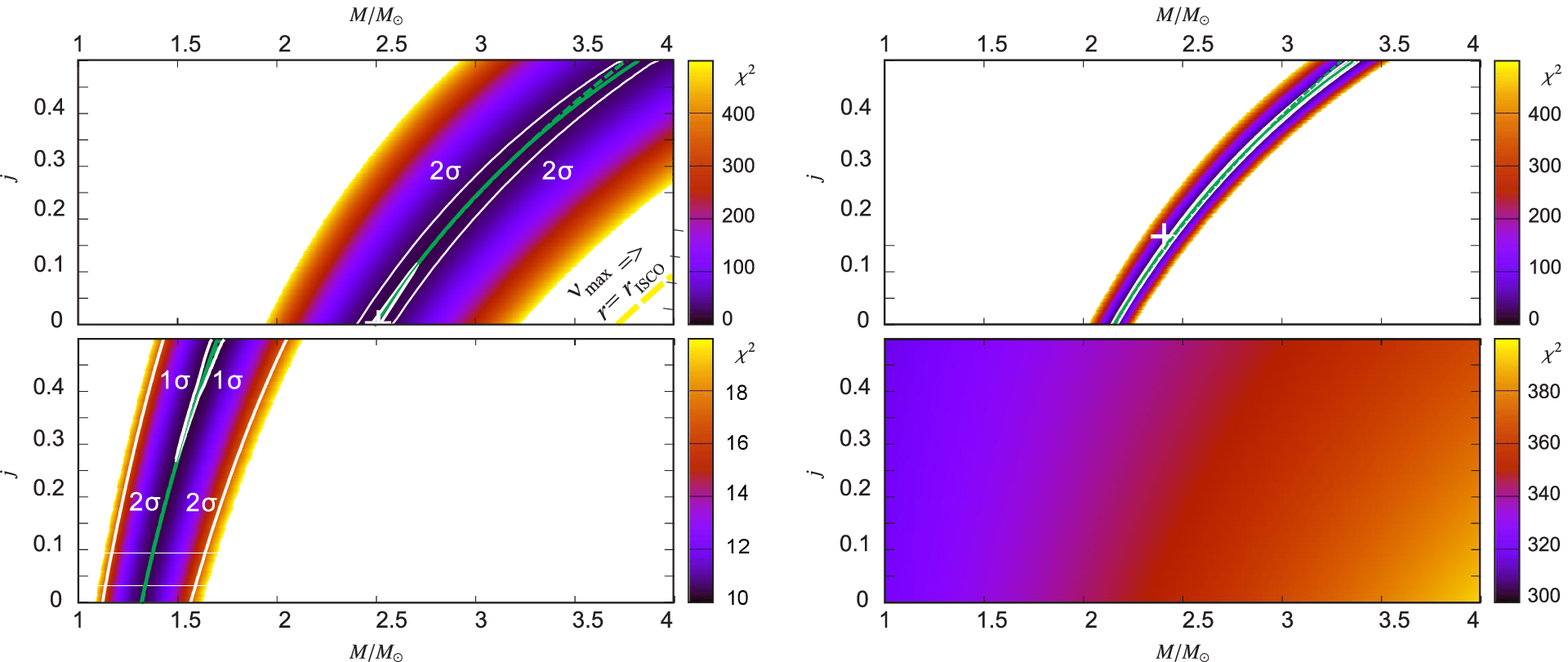}
\end{center}
\end{minipage}
\caption{ Same as Figure~\ref{figure:stella}(b), but for the other models.  {In 4U~1636--53  $\xi=4$ is assumed for the RP1 and RP2 models, $\xi=4.6$ for the WD model, and $\xi=2.5$ for the TD model.  For the TD model  the ISCO estimate on the upper limits on $M$ and $j$ from the highest observed QPO frequency in 4U~1636--53 is not included since the model does not associate this frequency to the ISCO but to the radius where the term $\nuK(r)+\nur(r)$ reaches its maximum.}}  
\label{figure:maps}
\end{figure*}
%---------------------------------------------------------

%--------------------------------------------------------------
\subsubsection{Results for the RP2 Model}
%--------------------------------------------------------------

As in the previous case, the frequencies predicted by the RP2 model are  very similar to those of the RP model for a slowly rotating NS. The lower observable QPO frequency is commonly defined by Equation~(\ref{equation:periastron}). The upper observable QPO frequency differs from the RP model and reads
%--------------------------------------------------------------
\begin{equation}
\quad \nuU^\mathrm{~\R\P\two}=2\nuK-\nuv.
\end{equation}
%--------------------------------------------------------------
However, in the Schwarzschild limit $j=0$, $\nuv=\nuK$ and the expression for the upper observable QPO frequency $\nuU=\nuK$ is common for all the three models RP, RP1, and RP2. For $j=0$, therefore, the frequency relations implied by these models merge (although the expected mechanisms generating QPOs are different). Thus we can write
%--------------------------------------------------------------
\begin{eqnarray}
\label{equation:RP12:fits2}
M_0^\mathrm{~\R\P\two}=M_0^\mathrm{~\R\P\one}=M_0^\mathrm{~\R\P}\,,
\end{eqnarray}
%--------------------------------------------------------------
where $M_0^\mathrm{~\R\P}$ is given in  Equation (\ref{equation:RP:fit:1636}) for 4U~1636--53 and Equation  (\ref{equation:RP:fit:circinus}) for Circinus X-1. {For 4U~1636--53, the quality of the fits is again not much different between $j=0$ and $j\neq0$ and  the same conclusions are valid  on the possible unknown systematic uncertainty as in the case of  {the RP1 and RP2} models.}

The $\chi^2$ behavior resulting from fitting the data points for free $M$ and $j$ is depicted in the form of color-coded maps in Figure~\ref{figure:maps}(b). These $\chi^2$ maps again clearly reveal $M-j$ degeneracy qualitatively similar to that in the case of RP and RP1 models. The best $\chi^2$ for a fixed $M$ ($M-j$ relation) is in each case denoted by the dashed green line.  {The corresponding approximate relations in the form $M=M_0\times[1+k(j+j^2)]$ read}
%--------------------------------------------------------------
\begin{eqnarray}
\label{equation:RP2:Mjfit:1636}
&&M=1.78[\pm {0.03}]M_\sun\times[1+0.98(j+j^2)]\nonumber\\
&&\mathrm{in~4U~1636-53}\\
\mathrm{and}&&\nonumber\\
\label{equation:RP2:Mjfit:circinus}
&&M=2.19[\pm {0.3}]M_\sun\times[1+ {0.65}(j+j^2)]\nonumber\\
&&\mathrm{in~Circinus~X-1}.
\end{eqnarray}
%--------------------------------------------------------------

%--------------------------------------------------------------
\subsection{Results for the WD Model}
%--------------------------------------------------------------

 {Considering $j=0$ for fitting the data of 4U~1636--53 we find a narrow $\chi^2$ minimum for $M_0\sim2.5M_\sun$ but its absolute value is  somewhat higher than in the case of the RP model, $\chi^2\doteq {450/21} \mathrm{dof}$.} {Moreover, there is also no sufficient improvement along the whole  given range of mass even up to the upper limit of $j$. Thus, we can again only speculate that there is an unknown systematic uncertainty.}
{The $\chi^2$ of the best fit for $j=0$ drops to an acceptable value $\chi^2=1 \mathrm{dof}$ for $\xi\doteq4.6$. The related mass corresponding to the best fit then reads $M_0=2.49[ {\pm0.1}]M_\sun$.} 

{In analogy to the RP model, the best match to the data of Circinus~X-1 for $j=0$ reveals an acceptable value of $\chi^2\doteq {10.6/10} \mathrm{dof}$. In summary, we can write the quantities $M_0$ for both sources as}
%--------------------------------------------------------------
\begin{eqnarray}
\label{equation:WD:fit:1636}
&&M_0=2.49[ {\pm0.1}]M_\sun\nonumber\\
&&\mathrm{in~4U~1636-53}\quad(\chi^2=1 \mathrm{dof}\Leftrightarrow\xi=4.6)\,\\
\mathrm{and}&&\nonumber\\
\label{equation:WD:fit:circinus}
&&M_0=1.31[ {+0.3,-0.2}]M_\sun\nonumber\\
&&\mathrm{in~Circinus~X-1}\quad(\chi^2= {10.6/10} \mathrm{dof}).
\end{eqnarray}
%--------------------------------------------------------------

The $\chi^2$ behavior resulting from fitting the data points for free $M$ and $j$ that again exhibits the $M-j$ degeneracy is depicted in Figure~\ref{figure:maps}(c).  The exact $M-j$ relations  in this figure are denoted by the dashed green lines.  {Their approximations in the form $M=M_0\times[1+k(j+j^2)]$ are, as in the previous cases, marked by the continuous green lines and read}
%--------------------------------------------------------------
\begin{eqnarray}
\label{equation:WD:Mjfit:1636}
&&M=2.49[ {\pm{0.1}}]M_\sun\times[1+0.68(j+j^2)]\nonumber\\
&&\mathrm{in~4U~1636-53}\\
\mathrm{and}&&\nonumber\\
\label{equation:WD:Mjfit:circinus}
&&M= {1.31}[ {+0.3,-0.2}]M_\sun\times[1+ {0.4}(j+j^2)]\nonumber\\
&&\mathrm{in~Circinus~X-1}.
\end{eqnarray}
%--------------------------------------------------------------

%---------------------------------------------------------
%                            FIGURE
%---------------------------------------------------------

\begin{figure*}[t!]

\begin{minipage}{1\hsize}
\begin{center}
(a)\hfill ~~~~(b) \hfill~
\\
\includegraphics[width=.95\textwidth]{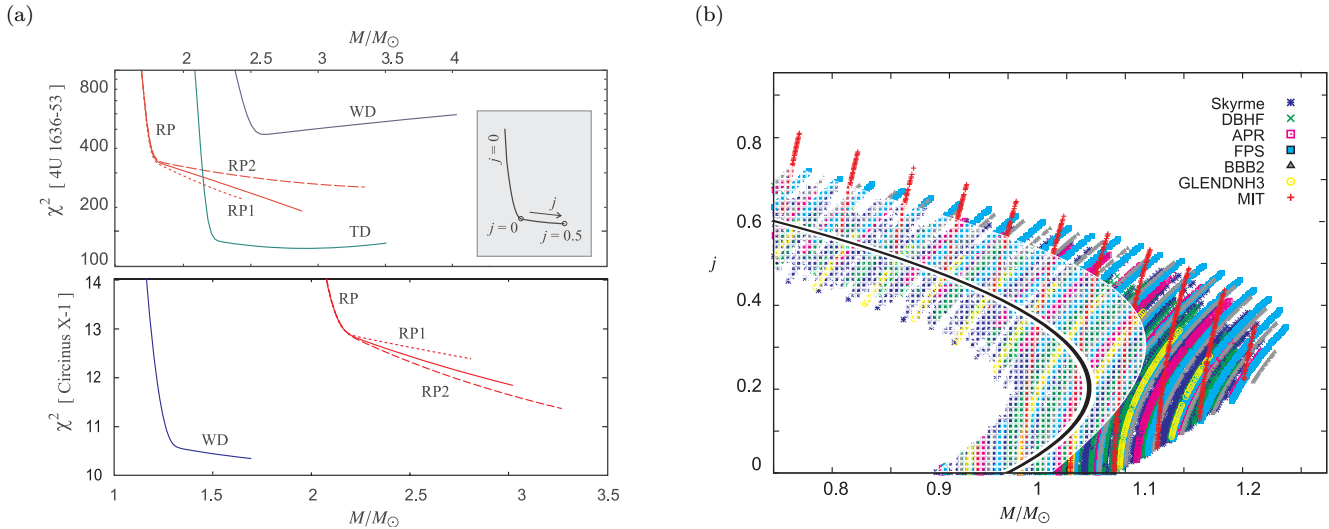}
\end{center}
\end{minipage}
\caption{ (a) Profiles of the lowest $\chi^2$ for a given $M$ plotted for various models. As in the previous figures, the case of 4U~1636--53 is shown in the top panel and Circinus X-1 in the bottom panel. The schematic drawing in the inset indicates the relation between the $\chi^2$ behavior and $j$ common to all the plotted curves. (b) The mass--angular-momentum combinations allowed by the ER model. The color symbols indicate different equations of state \citep[after][see these papers for details]{urb-etal:2010:aca, urb-etal:2010}. The lightened subset of these symbols is compatible with the 4U~1636--53 data. The black line denotes its quadratic approximation (Equation(\ref{equation:ER1})).}  
\label{figure:urbanec}
\end{figure*}
%----------------------------------------

%-------------------------------------------------------------- 
\subsection{Results for the TD Model}
%--------------------------------------------------------------

%--------------------------------------------------------
%--------------------------------------------------------
%														TABLE
%--------------------------------------------------------
%--------------------------------------------------------

\begin{deluxetable*}{lrrrrr}

\tablecaption{{The main definition relations for the  models considered and the mass--angular-momentum relations found for 4U~1636--53 and Circinus~X-1. }\label{table:1}}

\tablehead{
Model~~~~~~~~~ &  \multicolumn{3}{c}{atoll source 4U 1636-53} &  \multicolumn{2}{c}{Z-source Circinus X-1}\\
&  $\chi^2/\mathrm{dof}\sim$ &$\xi\sim$&  {{\Large{$\phantom{\frac{M}{M_\sun}}$}}} {{{$\left(M_0/M_\sun\right)\times f (j)$}}}~~~~~~~~& 
$\chi^2/\mathrm{dof}\sim$ &  {{\Large{$\phantom{\frac{M}{M_\sun}}$}}}{{{$\left(M_0/M_\sun\right)\times f (j)$}}}~~~~~~~~
}

\startdata
RP & & & & &\\
$\nuL = \nuK - \nur$, &
${16}$&
4.0&
 $1.78[\pm{0.03}]\times[1+0.7({j}+{j}^2)]$&
${{1.3}}$&
$ 2.2[\pm{0.3}]{}\times[1+0.5({j}+{j}^2)]$\\
$\nuU = \nuK$ & & &  & & \\
\\
TD & & & & &\\
$\nuL = \nuK$, &
${{7}}$&
2.5&
${2.15}[{\pm0.02}]~\times[1+0.7({j}+{j}^2)]$&
{{30}} &
 X$\phantom{^{**}}$\\
$\nuU = \nuK + \nur$& & & & &\\
\\
WD & & & &  &\\
$\nuL = 2(\nuK - \nur)$, &
${{21}}$&
4.6&
${2.49}[\,{\pm0.1}]\times[1+0.7({j}+{j}^2)]$&
${{1.1}}$& $ 1.3$\tablenotemark{a}\\
$\nuU = 2\nuK - \nur$ & & & & & \\
\\
RP1 & & & & &\\
$\nuL = \nuK - \nur$, &
${{16}}$&
4.0&
$ 1.78[{\pm{0.03}}]\times[1+0.5({j}+{j}^2)]$&
${{1.3}}$&
$2.2[\pm{0.3}]{}\times[1+0.4({j}+{j}^2)]$\\
$\nuU = \nuv$ & & & & &\\
\\
RP2 & & & & &\\
$\nuL = \nuK - \nur$, &
${{16}}$&
4.0&
$1.78[\pm{0.03}]\times[1+1.0({j}+{j}^2)]$&
${{1.3}}$&
$2.2[\pm{0.3}]{}\times[1+0.7({j}+{j}^2)]$\\
$\nuU = 2\nuK - \nuv$ & & & & &\\
\\
ER & & & & &\\
$\nu_L = \nur + \Delta\nuL$\tablenotemark{b}, &
${3}$&
1.7&
$ 0.95[\pm0.1]{}\times[1+0.8{j}-2{j}^2]$&
${1.5}$&
{$3.5[\pm0.3]{}\times[1+1.9(j+j^2)]$\tablenotemark{c}} \\
$\nu_{{U}} = \nuv + \Delta\nuU$& & & & &
\enddata

\tablecomments{Symbols $\nuK,~\nur,$ and $\nuv$ denote the orbital Keplerian, radial epicyclic, and vertical epicyclic frequencies (see Appendix~\ref{section:appendix:relations} for the explicit terms). For both sources, except for the ER model, the errors in the estimated mass corresponds to the $2\sigma$ confidence level. For the ER model, the errors are given by the scatter in the estimated resonant eigenfrequencies \citep[see][]{urb-etal:2010}.}

\tablenotetext{a}{The mass--angular-momentum relation that we found reads $M=1.3{[+0.3,-0.2]}M_\sun\times[1+0.4({j}+{j}^2)]$. Due to the low $M_0$, the $M(j)$ dependence cannot be taken seriously (see Section~\ref{section:conclusions} for a comment on this).}
\tablenotetext{b}{See Section~\ref{section:ER:model} for details.}
\tablenotetext{c}{The possibility that the observed frequencies are the combinational frequencies is taken into account.}

\end{deluxetable*}
%--------------------------------------
%--------------------------------------

 {Considering $j=0$ for fitting the data of 4U~1636--53 we find a narrow $\chi^2$ minimum for $M_0\sim{2.15}M_\sun$ {while its value $\chi^2\doteq {137/21}\mathrm{dof}$ is again unacceptable, although it is approximately $2\times$ lower than in the case of the RP model}. Moreover, there is also no sufficient improvement along the whole  given range of mass, even up to the upper limit of $j$. Thus, again we can only speculate that there is an unknown systematic uncertainty.}
{The $\chi^2$ of the best fit for $j=0$ drops to the acceptable value $\chi^2=1\mathrm{dof}$ for $\xi\doteq2.5$. The related mass corresponding to the best fit then reads: 
%--------------------------------------------------------------
\begin{equation}
M_0 = 2.15[\pm {0.02}]M_\sun\quad(\chi^2=1\mathrm{dof}\Leftrightarrow\xi=2.5)\,.
\end{equation}
%--------------------------------------------------------------
For the Circinus X-1 data we find no clear $\chi^2$ minimum. It is roughly $\chi^2\sim300/10\mathrm{dof}$ along the  interval of mass considered and $\chi^2$ is only slowly decreasing with $M$ decreasing (or $j$ increasing).

Color-coded maps of $\chi^2$ resulting for free $M$ and $j$ are shown in Figure~\ref{figure:maps}(d). In the case of 4U 1636--53 there is clearly an $M-j$ degeneracy. The $M-j$ relation is well approximated in the form $M=M_0\times[1+k(j+j^2)]$ as
%--------------------------------------------------------------
\begin{eqnarray}
\label{equation:TD:Mjfit:1636}
&&M=2.15[ {\pm0.02}]M_\sun\times[1+ {0.71}(j+j^2)].
\end{eqnarray}
%--------------------------------------------------------------

On the other hand, the $\chi^2$ distribution for Circinus~X-1 is rather flat, exhibiting roughly $\chi^2\sim300/10\mathrm{dof}$, whereas it slightly decreases for decreasing $M$ and increasing $j$. 

For the case of 4U~1636--53 the detailed profile of $\chi^2$ along the relation (\ref{equation:TD:Mjfit:1636}) is shown and compared to the RP, RP1, RP2, and WD models in Figure~\ref{figure:urbanec}(a). In the same figure we also show an analogous comparison for Circinus X-1. The absence of an $M-j$ relation and behavior of $\chi^2$ for the TD model in the case of Circinus X-1 is then discussed in Section~\ref{section:conclusions}.

%--------------------------------------------------------------
\subsection{Results for the ER Model}
\label{section:ER:model}
%--------------------------------------------------------------

Adopting the assumption that the observed frequencies are nearly equal to the resonant eigenfrequencies, $\nuU=\nuv(r)$ and $\nuL=\nur(r)$, the ER model does not fit the NS data \citep[e.g.,][]{bel-etal:2005,urb-etal:2010,lin-etal:2010}. A somewhat more complicated case in which this assumption is not fulfilled has been recently elaborated by \cite{urb-etal:2010}, who assumed data of 12 NS sources,  including 4U~1636-53. They investigated the suggestion made by \cite{abr-etal:2005a,abr-etal:2005b} that the resonant eigenfrequencies in 12 NS sources roughly read $\nuL^{\,0}=$~600Hz versus $\nuU^{\,0}=$~900Hz and the observed correlations follow from the resonant corrections to the eigenfrequencies, $\nuL=\nuL^{\,0}+\Delta\nuL$ versus $\nuU=\nuU^{\,0}+\Delta\nuU$. In this concept  {the resonance occurs at the fixed radius $r_{3:2}$ and} the data of the individual sources are expected as a linear correlation. Intersection of this  correlation with the $\nuU/\nuL=3/2$ relation gives the resonant eigenfrequencies since it is expected that $\Delta\nuL=\Delta\nuU=0$ when $R=3/2$. More details and references to the model can be found in \cite{urb-etal:2010}.

For the sake of the comparison with the RP and other models examined here, we plot Figure~\ref{figure:urbanec}(b) based on the results of \cite{urb-etal:2010}. The figure displays combinations of mass and angular momentum required by the model. The color-coded symbols indicate solutions for different equations of state (EoS). We denote the subset of these solutions compatible with the data of 4U~1636--53 by lighter symbols. The determination of this subset comes from the fit of 4U~1636--53 data by a straight line ($\chi^2=37/20/\mathrm{dof}$). It is clear from the figure that, as in the previous cases,  for the ER model there is a preferred mass--angular-momentum relation. In contrast to the other  models examined, it tends to a positive correlation between $M$ and $j$ only for low values of the angular momentum, $j\lesssim0.2$, while for a higher $j$ the required mass decreases with increasing $j$. This trend is connected to a high influence of the NS quadrupole momentum and large deviation from the Kerr geometry that arise for the low-mass NS configurations \citep[see][for details]{urb-etal:2010}. We find that the  mass--angular-momentum relation implied by the ER model for 4U~1636--53 can be approximated by a quadratic term roughly as (black curve in Figure~\ref{figure:urbanec}(b))
%--------------------------------------------------------------
\begin{equation}
\label{equation:ER1}
M=0.95{M_{\odot}}\times\left[1+0.8{j}-2{j}^2\right]\pm{10\%}\,.
\end{equation}
%--------------------------------------------------------------
For Circinus~X-1, the observed frequency ratio is far away from $R=3/2$ and the  ER model  assumed above cannot fit the Circinus X-1 data without additional assumptions. The high frequency ratio can be reproduced only if the resonant combination frequencies are taken into account \citep[e.g.,][]{tor-etal:2006}. In such a case, the lower observed QPO frequency would correspond to a difference between the resonant eigenfrequencies having values about (300Hz,~200Hz), i.e., approximately 3$\times$ less than the typical  twin peak QPO frequencies observed in 4U~1636--53. The related non-rotating mass would then be approximately 3$\times$ higher than that corresponding to 4U~1636--53, i.e., $M_0\sim3M_{\odot}$. The related fit of the Circinus X-1 data by a straight line has $\chi^2=16/10 \mathrm{dof}$. Taking  into account the change in eigenfrequencies due to the NS angular momentum and assuming the Kerr spacetime with $j<0.5$, we can express the formula for the mass of Circinus X-1 implied by the ER model approximately as
%--------------------------------------------------------------
\begin{equation}
\label{equation:ER2}
M=3{M_{\odot}}\times\left[1+1.9\left(j+j^2\right)\right]\pm{10\%}\,.
\end{equation}
%--------------------------------------------------------------

While for 4U~1636--53 the mass decreases with increasing $j$  (Equation (\ref{equation:ER1})), for Circinus X-1 the trend is opposite. This behavior is associated with the choice of the spacetime geometry. The low mass $M_0\sim1M_\sun$ inferred from the model for 4U~1636--53 implies high deviations from the Kerr geometry due to the NS oblateness \citep[][]{ urb-etal:2010:aca, urb-etal:2010}. In such situation orbital frequencies can decrease with increasing $j$.  For Circinus~X-1, the high mass $M_0=3M_\sun$ justifies the applicability of the  Kerr geometry chosen. For this geometry, the orbital frequencies must increase with increasing $j$ (provided that $j<1$). This issue is well illustrated by the behavior of ISCO frequencies in the right panel of Figure~3 in Paper~I.

%--------------------------------------------------------------
\section{Chi-squared Dichotomy and Corrections to the RP or Other Models}
%--------------------------------------------------------------
\label{section:dichotomy}

%--------------------- FIGURE ----------------------------------------------------
\begin{figure*}[t!]
\begin{minipage}{1\hsize}
\begin{center}
(a)\hfill (b) \hfill (c)\hfill{$\phantom{d}$}
\smallskip
\includegraphics[width=1\textwidth]{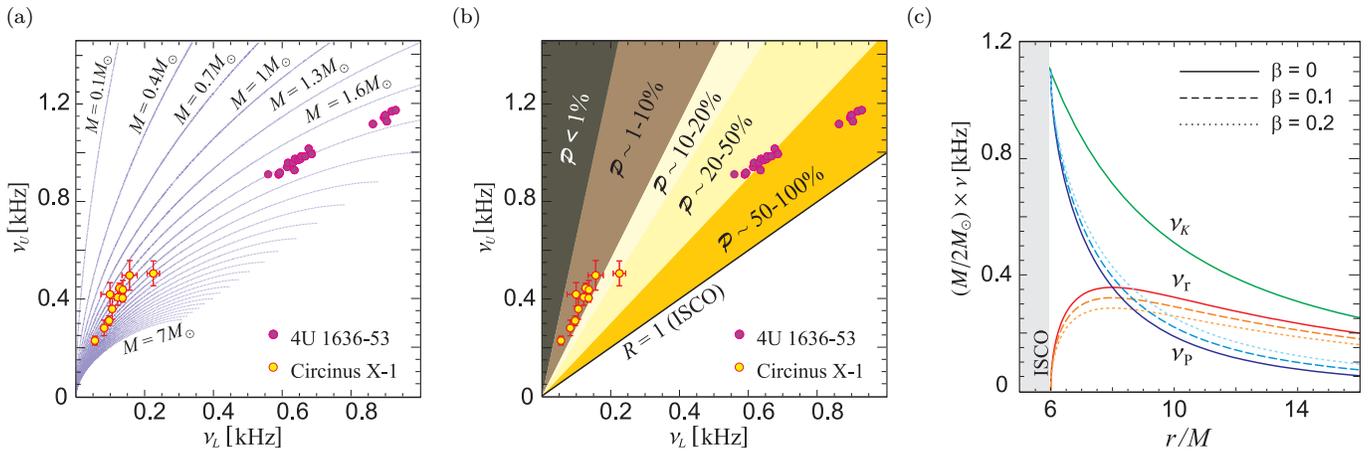}
\end{center}
\end{minipage}
\caption{ (a) Frequency relations predicted by the (geodesic) RP model for $j=0$ vs. data of 4U~1636-53 and Circinus X-1. (b) Quantity $P$ illustrating the variability of the predictive power of the RP model across the frequency plane. (c) Profiles of the orbital, radial epicyclic, and periastron frequencies of the perturbed circular motion. Solid curves correspond to the geodesic case ($\beta=0$). The dashed and dotted curves correspond to the case of non-geodesic radial oscillations ($\beta>0$).}
\label{figure:dichotomyf1}
\end{figure*}
%-------------------------------------------------------

It has been noticed by \cite{ste-vie:1999} and later by a number of other authors that data of sources with QPOs sampled mostly on low frequencies are better fitted by the RP model than data for sources with QPOs sampled mostly on high frequencies. Inspecting $\chi$-square maps (Figures~\ref{figure:stella} and \ref{figure:maps}) and Table~\ref{table:1}, we can see that the  comparison between Circinus X-1 (good $\chi^2$) and 4U 1636--53 (bad $\chi^2$) well demonstrates such a \qm{dichotomy}. The $\chi^2$ maps and profiles for the RP model are qualitatively similar for both 4U~1636--53 and Circinus X-1. Both sources also exhibit a decrease of $\chi^2$ with increasing $j$ (see Figure \ref{figure:urbanec}(a)).  {The $\chi^2$ {values} reached for 4U 1636--53 {are}, however, much worse than {those} in the case of Circinus X-1 ($\approx$10 versus 1 dof), and {their} spread with $M$ is much narrower.}  Moreover, we find that a similar dichotomy also arises for all the other  models  considered assuming that the observed  twin peak QPO frequency correlation arises directly from a correlation between characteristic frequencies of the orbital motion. Below we briefly discuss the relation between this dichotomy, the predictive power of the model, and possible non-geodesic corrections.  {We restrict our attention mostly to the RP model but argue that there is a straightforward generalization to the other models}.

%--------------------------------------------------------------
\subsection{Data versus Predictive Power of the RP Model}
%--------------------------------------------------------------
\label{section:predictive:power}

%One should be aware of the fact that the predictive power of the RP model strongly depends on the QPO excitation radius $r$.
Figure \ref{figure:dichotomyf1}(a) shows the frequency relations predicted by the RP model for a non-rotating NS and several values of mass $M$. These curves run from the common point $[\nuL,~\nuU]=[$0Hz,\,0Hz$]$ corresponding to  infinite $r$. They terminate at specific points $[\nu_\mathrm{ISCO},\,\nu_\mathrm{ISCO}]$ corresponding to $r=r_\mathrm{ms}=r_\mathrm{ISCO}$. %
This behavior follows the fact that for low excitation radii close to ISCO, a certain change  in $M$ leads to a modification of the orbital frequency that is much higher than those for radii far away from ISCO. In other words, the  predictive power of the RP model is much weaker for radii far away from ISCO than for radii close to ISCO.

%In terms of the observed quantitities the RP model predictive power is related to the values of frequency ratio $R$ rather than to absolute values of frequency.
As recalled in Section~\ref{section:introduction}, in the RP model the radius $r$ is  proportional to $R$ \citep[e.g.,][]{tor-etal:2008a}. Because of this, the predictive power of the RP model is strongly decreasing with increasing $R$. In Appendix~\ref{section:appendix:PP}  we discuss this in terms of the quantity $\mathcal{P}\propto R^{-3}$ determining the squared distance $ds$$^2$ measured in the frequency plane between data points related to different masses. This quantity has a direct impact on the spread of $\chi^2$. For a certain variation of the mass, $\delta\equiv \Delta M/M$, it is 
%--------------------------------------------------------------
\begin{equation}
{d}s^2\propto\frac{\delta^2}{(1+\delta)^2}\mathcal{P}\,.
%\mathrm{d}s^2\propto\frac{\delta^2}{(1+\delta)^2}\frac{1}{R^{3}}\,.
\end{equation}
%--------------------------------------------------------------
Detailed formulae are given in Equations~(\ref{equation:A:PP1}) and (\ref{equation:A:PP2}). Figure~\ref{figure:dichotomyf1}(b) shows behavior of $\mathcal{P}$ in the frequency plane.

Taking into account the data points included in Figures \ref{figure:dichotomyf1}(a) and (b) and the behavior of $\mathcal{P}$ we can deduce that the difference in the spread of $\chi^2$ in 4U~1636--53 and Circinus~X-1, as well as the very different values of the $\chi^2$ minima in these sources, can be related  to both the size of the error bars (affected by a low significance of kHz QPOs on low frequencies) and the location of data points. In Circinus X-1, the data points lie in the region of relatively low frequencies related to high $R$. For these, the predictive power of the model is low, since the curves $\nuU(\nuL)$ expected for various parameters $M$ and $j$ converge. On the other hand, in 4U 1636--53, the data points lie in the region of relatively high frequencies related to low $R$. These correspond to the strong gravity zone where  different correlations are much more distinguished and the predictive power of the model is high.  {Similar consideration is also valid for several other models that predict frequency curves converging at low $R$. Clearly, from Figures \ref{figure:stella} and \ref{figure:maps} we can see that the uncertainties of the inferred mass expressed at $2\sigma$ confidence levels in 4U~1636--53 are $\sim10\!-\!20\times$ smaller compared to Circinus X-1 for each of the RP, RP1, RP2, and WD models.} 

%--------------------------------------------------------------
\subsection{Toy Non-geodesic Modification of the RP Model}
%--------------------------------------------------------------

%--------------------- FIGURE ----------------------------------------------------
\begin{figure*}[t!]
\begin{minipage}{1\hsize}
\begin{center}
(a)\hfill ~~~(b) \hfill (c)\hfill{$\phantom{d}$}
\medskip

\includegraphics[width=1\textwidth]{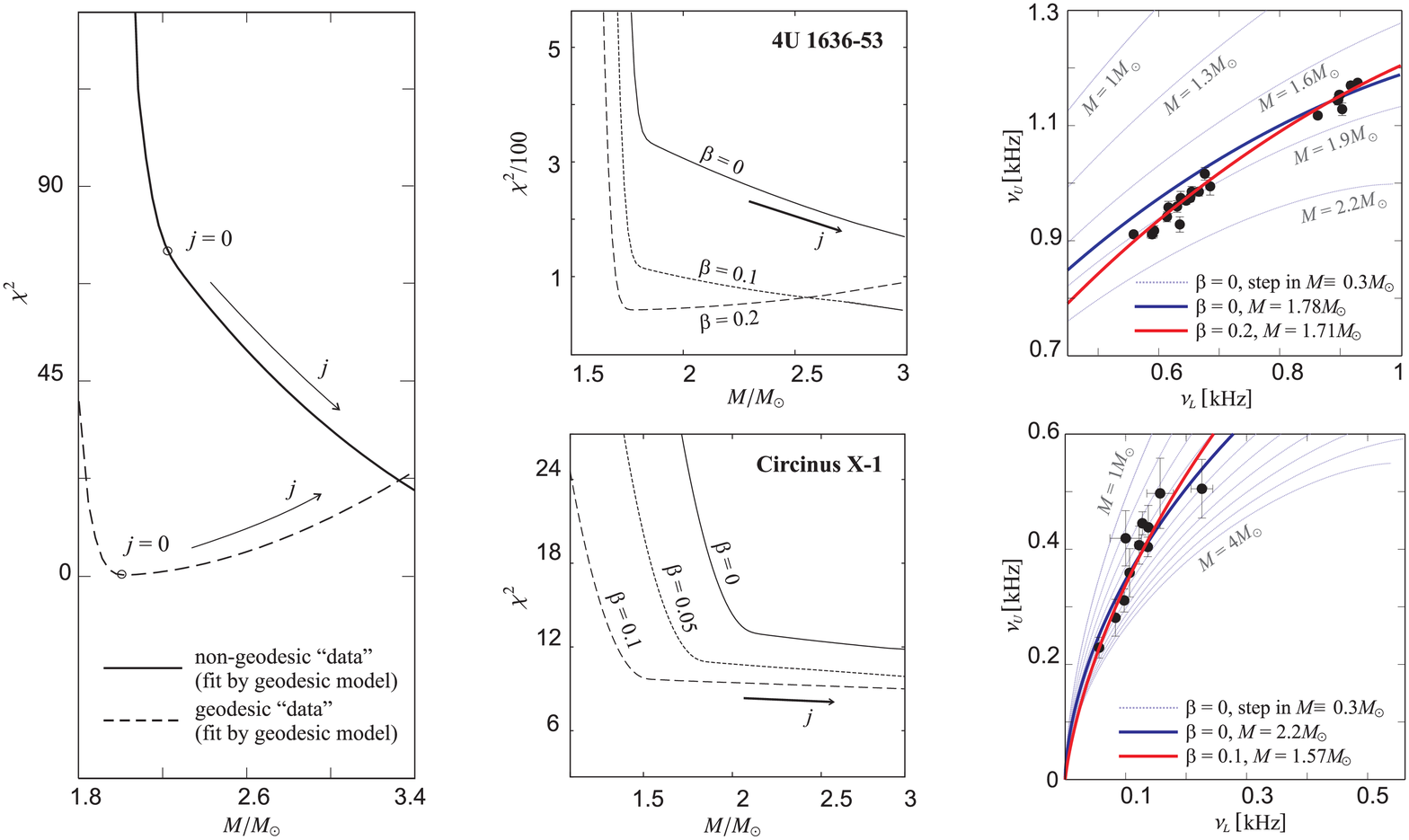}
\end{center}
\end{minipage}
\caption{ (a) Profile of the best $\chi2$ for a fixed $M$ calculated
when the simulated data are matched by the geodesic RP model. The continuous line is plotted for $M=2M_\sun$, $j=0$, and
$\beta=0.1$. The dashed line is plotted for $\beta=0$. The arrows indicate increasing j. (b) Profiles of the best $\chi2$ for a fixed $M$ 
in the case when Equation (\ref{equation:modification}) is assumed
for fitting of the real data. The arrows in each panel indicate increasing j. The vertical arrow denotes the
improvement $\Delta  \chi^2$.
(c) Comparison of the geodesic ($\beta=0$, thick blue line) and non-geodesic ($\beta>0$, red line) fits is included in the \qm{zoom} from Figure~\ref{figure:dichotomyf1}(a). The top panel is plotted for 4U~1636-53 while the bottom panel is plotted for Circinus X-1. Both panels have the same scaling of the axes.\\ 
}
\label{figure:dichotomyf2}
\end{figure*}
%---------------------------------------- 

Based on the above findings, we can speculate that the same systematic deviation from the  {particular  model considered} may be involved in both sources. We justify this speculation using an arbitrary example of a toy non-geodesic version of the RP model. We attempt to use a modification that would mimic the behavior of real data. In the vicinity of the inner edge of an accretion disk it is natural to expect a modification of the radial epicyclic frequency rather than a modification of the Keplerian frequency. The orbital motion in this region is highly sensitive to radial perturbations and even very small deviations from the geodesic idealization can strongly affect the radial oscillations \citep[see in this context][]{stu-etal:2011}. In our example we therefore assume that the frequency of the hot-spot radial oscillations is somewhat lowered due to pressure or magnetic field effects \citep[e.g.,][]{str-sra:2009,bak-etal:2010,bak-etal:2012}. For simplicity, we postulate that the effective frequency of the radial oscillations is
%--------------------------------------------------------------
\begin{equation}
\tilde{\nur}=\nur(1-\beta),
\end{equation}
%--------------------------------------------------------------
where $\beta$ is a small constant. The related lower QPO frequency actually observed is then given by
%--------------------------------------------------------------
\begin{equation}
\label{equation:modification}
\tilde{\nul}=\nul + \beta\left(\nuu  - \nul\right),
\end{equation}
%--------------------------------------------------------------
where  $\nul(\nuU)$ is the frequency relation of the geodesic RP model given in Equation~(\ref{equation:A:RP}). Assuming Equation~(\ref{equation:modification}), $\beta=0.1$, $j=0$, and $M=2M_\sun$ we produce 20 data points uniformly distributed  along the frequency correlation. We then fit the simulated data by the geodesic model.  Figure~\ref{figure:dichotomyf2}(a) shows the resulting $\chi^2$ profile calculated in the same way as those in Figure~\ref{figure:urbanec}(a). Clearly, $\chi^2$ decreases with growing $j$ similarly to the results obtained for real data points in both  sources discussed. For comparison, we also present the fit of data simulated for $\beta=0$, where, in contrast, $\chi^2$ increases with growing $j$. Having this boost we use Equation (\ref{equation:modification}) for the fitting of the real data points. The resulting "best $\chi^2$" improves for both sources, although in the case of Circinus X-1 the improvement is only marginal. More specifically, for 4U~1636--53 the best $\chi^2$ improves up to $\beta\sim0.2$ with $\Delta \chi^2\sim300$, while for Circinus X-1 it improves up to $\beta\sim0.1$ with $\Delta \chi^2\sim4$. The representative $\chi^2$ profiles are illustrated in Figure~\ref{figure:dichotomyf2}(b), which also shows the related impact on mass restrictions.  The strong improvement in 4U~1636--53 data corresponds to only a marginal effect on the mass restriction ($\Delta M \lesssim0.1M_\sun$). On the other hand, the small improvement of $\chi^2$ in Circinus X-1 causes a large modification of the mass restriction ($\Delta M \sim0.6M_\sun$). The related fits to the data are shown in  Figure~\ref{figure:dichotomyf2}(c). 

The toy model (\ref{equation:modification}) naturally does not represent an elaborate attempt to describe the QPO mechanisms,  but it  demonstrates well that, in spite of the good quality of  fit, in both 4U 1636--53 and Circinus X-1 sources, the same physical correction to the RP model could be involved. A similar consideration should also be  valid for several other  models discussed. {In this context, we note that sophisticated implementations of non-geodesic corrections have been developed in the past within the framework of various models of accretion flow dynamics and QPOs \citep[see, e.g.,][and references therein]{wag-etal:1999,wag-etal:2001,kat:2001,alp-psa:2008}. We also note that some corrections to the orbital frequencies can arise directly due to corrections to the Kerr or Hartle--Thorne (HT) spacetimes that we assume here \citep[see, e.g.,][]{kot-etal:2008,psa-etal:2008,stu-kot:2009,joh-psa:2011}.}

%---------------------------------------------------------
\begin{figure*}[t]
(a)\hfill ~~~~~~~~(b) \hfill~
\smallskip

\begin{minipage}{1\hsize}
\begin{center}
\includegraphics[width=.8\textwidth]{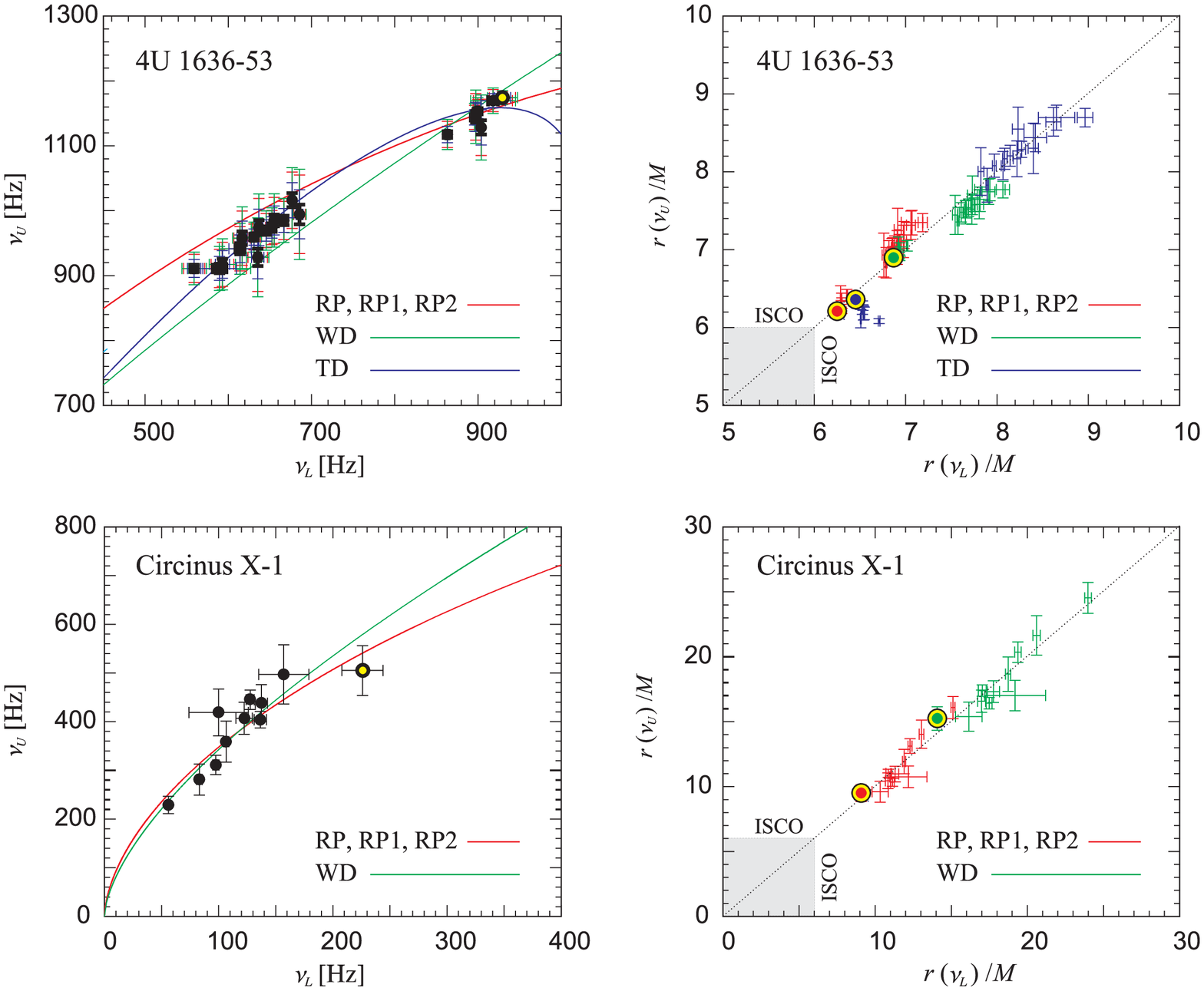}
\end{center}
\end{minipage}
\caption{  {Best fits to the data by individual models for $j=0$. (a)  Frequency relations. Error bars corresponding to $\xi=4$ for RP models, $\xi=4.6$ for WD model, and $\xi=2.5$ for TD model are color-coded. The pair of the highest twin peak QPO frequencies observed in the source is marked by a yellow circle. (b) The QPO excitation radii inferred from the data and each of the fits that are shown in the panel (a). The color-coded circles correspond to the highest observed twin peak QPO frequencies. The TD model is included for 4U~1636-53 only because it does not match the data of Circinus~X-1 (see Section~\ref{section:conclusions} for a discussion).}}  
\label{figure:fits}
\end{figure*}
%----------------------------------------

%--------------------------------------------------------------
\section{Discussion and Conclusions}
\label{section:conclusions}
%--------------------------------------------------------------

%\issues{TBD: SUMMARY on the new Figure (fits for j=0 and radii implied by QPOs) drawn for RP, WD, TD and ER models.}

Except the TD model applied to Circinus~X-1 data, all applications of the  models examined to the 4U~1636--53 and Circinus X-1 data result in the preferred mass--angular-momentum relations. These are summarized in Table~\ref{table:1}.

Comparing the $\chi^2$ map of  the TD model and Circinus~X-1 (Figure~\ref{figure:maps}(d)) to the other $\chi^2$ maps we can see that it is very different with its flat $\chi^2$ behavior. Moreover, the TD model is the only model of those considered here giving very bad $\chi^2$ for Circinus~X-1 ($\chi^2\sim$300/10 dof versus $\chi^2\sim$10/10 dof for the other models). This can be well understood in terms of the frequency ratio $R$ implied by the model. The TD model states 
%--------------------------------------
\begin{equation}
\nuL=\nuK\,,\quad\nuU=\nuK+\nur,
\end{equation}
%--------------------------------------
where $\nur\leq\nuK$. In more detail, $\nur$ vanishes at $r=r_\mathrm{ISCO}$ and, in a flat spacetime limit ($r=\infty$), $\nur=\nuK$. Consequently, the TD model allows only $R\in(1,~2)$. The Circinus~X-1 data, however,  reveal  values between $R\sim2.5$ and $R\sim4.5$ which is clearly higher than the Newtonian limit, $R=2$. This disfavors the TD model.
%\bigskip\bigskip

\subsection{ {Quality of Fits and Inferred Masses: Models with $\nu(r)$}}

{ Table~\ref{table:1} provides a summary of results of fits to the data for both sources by individual models. The comparison between fits by individual models is illustrated in Figure~\ref{figure:fits} which also indicates the inferred QPO excitation radii. Within the RP, RP1, RP2, and WD models, the quality of fits is rather comparable (bad for 4U~1636 and good for Circinus~X-1). The mass--angular-momentum relations are similar for the RP, RP1, and RP2 models while for the WD model they differ (see Table~\ref{table:1}). In more detail, the RP, RP1, and RP2 models require relatively similar masses for both sources, namely $M_0\sim1.8M_\sun$ for 4U~1636--53 versus $2.2M_\sun$ for Circinus X-1. On the other hand, the required masses differ quite a lot when the WD model is assumed. We then have $M_0\sim2.5M_\sun$ for 4U~1636--53 versus $1.3M_\sun$ for Circinus X-1.  {We note that the QPO excitation radii inferred for each model in 4U~1636--53 lie within the innermost part of the accretion disk. This is depicted in detail in Figure~\ref{figure:fits}(b) assuming a non-rotating star. We can see that the radii span the interval $r\in(6M-8M)$ for the RP model, $r\in(7M-8M)$ for the WD model, and the largest interval $r\in(6M-9M)$ for the TD model. On the contrary, the radii inferred in Circinus~X-1 are above $r=10M$, belonging to the interval $r\in(10M-16M)$ for the RP model and $r\in(15M-25M)$ for the WD model.}}

%--------------------- FIGURE ----------------------------------------------------
\begin{figure*}[t!]
\begin{minipage}{1\hsize}
\begin{center}
%a)\hfill $\phantom{b}$ \hfill $\quad\quad\quad$b) \hfill $\phantom{b}$ \hfill $\phantom{b}$ \hfill{$\phantom{d}$}
(a)\hfill (b) \hfill ~~~~~~~$\phantom{c}$\\

\includegraphics[width=.98\textwidth]{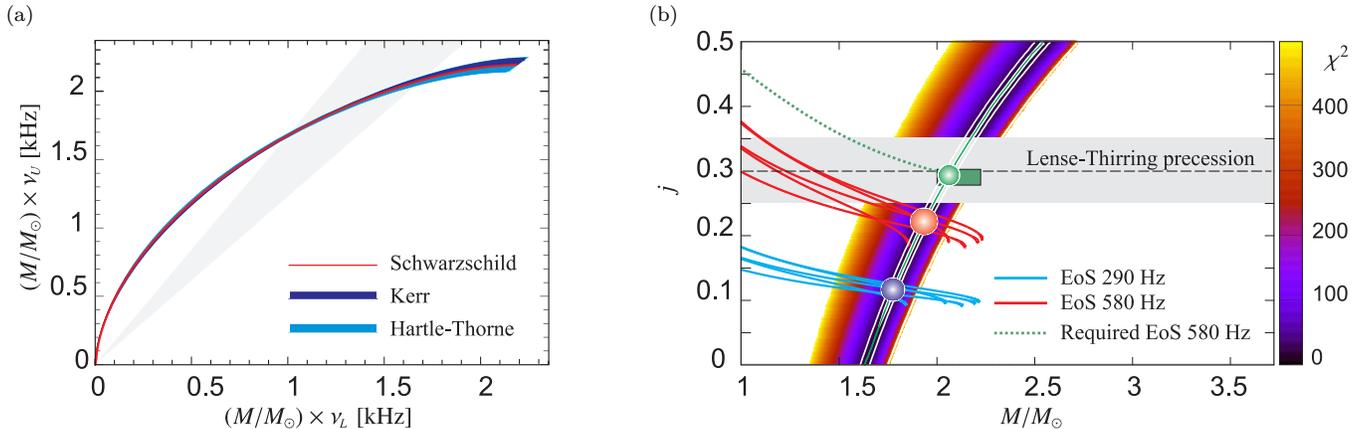}
\end{center}
\end{minipage}
\caption{ 
(a) The ambiguities of parameters of RP model frequency relations illustrated for the range $j\in(0,~0.3)$ and $\tilde{q}\in(1,~8)$. The red curve indicates the relation plotted for the mass $M_0$ in the Schwarzschild spacetime. The dark blue set of curves marked as \qm{Kerr} represent the degeneracy in the Kerr spacetimes given by Equation~(\ref{equation:kerr07}). The light blue set marked as \qm{Hartle-Thorne} includes curves resulting from the generalized degeneracy in HT spacetimes given by Equation~(\ref{equation:HT:32}). The shadow cone denotes the range of frequency ratio $R$ corresponding to the data of 4U~1636-53.
(b) Removing the $M-j$ degeneracy in the case of 4U~1636-53 and the RP model. The  $\chi^2$ map displayed is calculated for $\beta\neq0$ while the best fits correspond to $\beta=0.15-0.20$. The blue spot roughly indicates the combination of mass and spin restricted when the spin frequency 290Hz and several concrete equations of state are assumed. The red spot indicates the same but for the spin frequency 580Hz. The shaded region around the dashed horizontal line indicates the angular momentum $j=0.3\pm0.05$ which can be roughly expected when the Lense-Thirring precession is assumed. The green box corresponds to a detailed consideration of a few points in the 3D frequency space (see Appendix~\ref{section:appendix:RP} for details). }
\label{figure:EOSLTP}
\end{figure*}
%---------------------------------------- 

The above models have, along  with few others, recently been considered for 4U~1636--53 by \cite{lin-etal:2010}. They reported mass and angular momentum corresponding to $\chi^2$ minima for each of the models. The data points they investigated especially for this purpose come from a sophisticated, careful application of a so-called shift-add procedure over a whole set of the available $RXTE$ observations (see their paper for details and references). The data we use here for 4U~1636--53 come from the previously well-investigated individual continuous observations of the source \citep[see][]{bar-etal:2005a,bar-etal:2005b,tor:2009}. While the two sets of the applied data come from different methods, the values of mass and angular momentum reported by \cite{lin-etal:2010} agree with the mass--angular-momentum relations that we find here (see Figures \ref{figure:stella}(b) and \ref{figure:maps}).%\footnote{For the sake of completeness, we also searched for the mass--angular-momentum relation for datapoints used in \cite{lin-etal:2010}. The results agreed within errors with the mass-angular-momentum relations reported here and summarized in Table~\ref{table:1}.}
 One should note that, in contrast to $M-j$ relations, the single $M-j$ combination corresponding to the $\chi^2$ minimum of a given model is not very informative as the (bad) $\chi^2$ is comparable along a large range of mass. Moreover, in each case examined here the $\chi^2$ minima correspond only to the end of the  angular momentum interval considered since the quality of fit is a monotonic function of $j$. 
{Thus, we can conclude that the differences between the $M_0$ coefficients in Table $\ref{table:1}$ provide the main information about the differences between predictions of the individual QPO models.}

 {In relation to the quality of fits by the RP, RP1, RP2, and WD models, we {can also  note} that these models need some correction, as has also been noted by \cite{lin-etal:2010}. As demonstrated in Section~\ref{section:dichotomy},  differences in the $\chi^2$ behavior between low- and high-frequency sources can be related to the variability of the model's predictive power across the frequency plane. This variability naturally comes from the radial dependence of the characteristic frequencies of orbital motion. As a consequence, the restrictions to the models resulting from the observations of  low-frequency sources are weaker than those in case of high-frequency sources. A small required correction is then likely to be common to both classes of sources, which has been demonstrated using the non-geodesic modification of the RP model based on Equation~(\ref{equation:modification}).}

\subsubsection{ {Applicability of Results Based on the Spacetime Description Adopted}}

 {Both the poor quality of  fits to the data by geodesic models and the mass--angular-momentum relations associated with these models have been obtained assuming the Kerr spacetimes. This approximate description of the exterior of rotating NS neglects the NS oblateness. As argued in Paper~I, the uncertainty in NS oblateness causes only small inaccuracies in the modeling of kHz QPOs for the compact high-mass NSs. In the case of the WD model applied to the Circinus X-1 data, a consequent application of a more sophisticated approach is still needed. The Kerr approximation suggested in Paper~I is clearly not valid here due to low $M_0\sim1.3M_\sun$. Such a low mass can imply high deviations from the Kerr geometry due to the strong influence of the NS oblateness. In principle, the related mass--angular-momentum relation can be very different in this case from that qualitatively implied, e.g., by the RP model, and corrections to the quadrupole moment should be included in analogy to the ER model and 4U~1636--53. For the other applications of the WD, RP, RP1, and RP2 models reported here we can trust the $M-j$ trends following from the Kerr approximation since the inferred masses $M_0$ are rather high.}

 {We justify the applicability of our results in Appendix~\ref{section:appendix:HT}. In general, the differences between geodesic frequencies associated with Kerr spacetimes and those given for realistic NSs due to their oblateness are roughly of the same order as the corrections required to obtain a good match between the predicted and observed QPO frequencies {\citep[e.g.,][]{mor-ste:1999}}. We illustrate however, that these differences cannot improve the fits sufficiently for NSs with $j\lesssim0.3$ and $M\gtrsim1.4M_\sun$. We show that in HT spacetimes describing the exterior of oblate NSs there is a degeneracy not only between the NS mass and angular momentum but also between these quantities and the NS quadrupole moment $q$. Within such \qm{generalized degeneracy} the frequency curves predicted by QPO models scale with the quantities $M,~j$, and $q$ but the related qualitative change in their shape is only small. Thus, our results obtained for the Kerr spacetimes have a more general relevance except for the case of high values of $j$ (see Appendix~\ref{section:appendix:HT} for details).}

\subsubsection{ {Prospects of Eliminating the $M-j$ Degeneracy}}
\label{section:conclusions:prospects}

{The $M-j$ degeneracies implied by individual kHz QPO models can in principle be eliminated using angular momentum estimates independent of the kHz QPOs. In Appendix~\ref{section:appendix:RP} we subsequently focus on the RP model and discuss such possible elimination. Based on the X-ray burst observations of \cite{str-mar:2002} we assume that the rotational frequency (spin) of the NS in 4U~1636--53 is around 290Hz or 580Hz. Applying few concrete NS EoS we show that the modified RP model well matches these spins for $j\sim0.1$ or $j\sim0.2$. We also show that a further consideration of low-frequency QPOs and the Lense--Thirring precession mechanism within the model can be finally crucial for fixing the value of $j$ and challenging for application of the concrete EoS (see illustration in Figure~\ref{figure:EOSLTP}). We note that this issue as well as modeling of kHz QPO correlations for rapidly rotating NS require an additional detailed treatment.}

\subsection{ {Resonance between $m=0$ Axisymmetric Disk-oscillation Modes}}

Last but not least, we can draw conclusions about the  version of the ER model  examined  {assuming the fixed radius $r=r_{3:2}$}. It well fits the data of both the sources discussed here with a $\chi^2/$dof of the order of unity. The good fits, however, arise because the present model predicts a linear correlation which has  slope and intercept given by unspecified (free) parameters. One should also note that for Circinus~X-1 the model requires additional consideration of the resonant combinational frequencies. Moreover, application of the ER model leads to a questionably low mass for 4U 1636--53, $M\leq 1M_\sun$, while for Circinus~X-1 the implied mass is on the contrary questionably high, $M\geq3M_\sun$. All these along with the results of \cite{urb-etal:2010} suggest that if a resonance is involved in the process of generating the NS QPOs, modes other than those corresponding to the radial and vertical axisymmetric oscillations should be considered.

%---------------
\acknowledgments
We thank Marek Abramowicz, Wlodek Klu\'zniak, Milan \v{S}enk\'y\v{r}, and Yong-Feng Lin for discussions. We also thank to the anonymous referee for his/her comments and suggestions that helped greatly to improve the paper.  {This work has been supported by the Czech grants MSM 4781305903, LC 06014, GA\v{C}R 202/09/0772, and GA\v{C}R 209/12/P740. The authors further acknowledge the project CZ.1.07/2.3.00/20.0071 "Synergy" supporting the international collaboration of IF Opava and also the internal student grants of the Silesian University in Opava, SGS/1/2010 and SGS/2/2010.}
%--------------------------------

%$$s=d\quad s=\sqrt{2}d \quad \chi^2=1 \quad \chi^2=2$$

%-----------------------------------------------------------
\appendix               % APPENDIX 
%-----------------------------------------------------------

%------------------------------------------------------------------------
\section{Approximations, Formulas, and Expectations}
%------------------------------------------------------------------------

%--------------------------------------------------------
\subsection{{Relations for the Upper and Lower QPO Frequencies in the RP, TD, WD, RP1, and RP2 Models}}
%--------------------------------------------------------
\label{section:appendix:relations}

Formulas for the Keplerian, radial, and vertical epicyclic frequency were first derived by \cite{ali-gal:1981}. In a commonly used form \citep[e.g.,][]{tor-stu:2005} they read
%--------------------------------------------------------
\begin{equation}
\label{equation:frequencies}
\Omega_\K=\frac{\B}{j+{x}^{3/2}},\quad\nur=\Gamma\Omega_\K,\quad\nuv=\Delta\Omega_\K\,,
\end{equation}
%--------------------------------------------------------
where
\begin{equation}
\Gamma =\sqrt{\frac{-3 j^2+8 j \sqrt{{x}}+\left(-6+{x}\right) {x}}{{x}^2}}, 
\quad \Delta = \sqrt{1+\frac{j \left(3 j-4 \sqrt{x}\right)}{x^2}}\,,
\end{equation}
%--------------------------------------------------------
$x\equiv r/M$, and the "relativistic factor" $\B$ reads $\B \equiv c^3/(2\pi GM)$.

Relations defining the upper and lower QPO frequencies in terms of the orbital frequencies are given for each of the  models considered in the first column of Table~\ref{table:1}. For the RP model, one can easily solve these relations to arrive at an explicit formula which relates the upper and lower QPO frequencies in the units of Hertz as (Paper~I)
%--------------------------------------------------------------
\begin{equation}
\label{equation:A:RP}
\nuL = \nuU\left\{1 - \left[1 + \frac{8j\nuU}{\B - j\nuU} - 6\left(\frac{\nuU}{\B - j\nuU}\right)^{2/3} \\ 
- 3j^2\left(\frac{\nuU}{\B - j\nuU}\right)^{4/3} \right]^{1/2}\right\}\,.
\end{equation}

%--------------------------------------------------------------

A similar simple evaluation of the explicit relation between the two observed QPO frequencies is also possible for the TD model, where we find
%--------------------------------------------------------------
\begin{equation}
\label{equation:A:TD}
\nuU = \nuL\left\{1 + \left[1 + \frac{8j\nuL}{\B - j\nuL} - 6\left(\frac{\nuL}{\B - j\nuL}\right)^{2/3} - 3j^2\left(\frac{\nuL}{\B - j\nuL}\right)^{4/3} \right]^{1/2}\right\}\,.
\end{equation}
%--------------------------------------------------------------
An apparent \qm{asymmetry} between relations (\ref{equation:A:RP}) and (\ref{equation:A:TD}) arises from an analogical asymmetry in the model definition of the observable frequencies (see Table~\ref{table:1}). We note that in both models, one of the two observable frequencies simply equals to the Keplerian orbital frequency, which makes the evaluation of the explicit formula very straightforward.

For the WD, RP1, and RP2 models, the definition relations lead to high-order polynomial equations that relate the lower and upper QPO. In these cases we can give only parametric form relating $\nuU$ and $\nuL$. 
The upper and lower QPO frequencies for the WD model can be then expressed as 
%--------------------------------------------------------
\begin{eqnarray}
\label{equation:A:WD}
\nuU=2\left(1-\Gamma\right)\Omega_\K\,,\quad
\nuL=\left(2-\Gamma\right)\Omega_\K\,.
\end{eqnarray}
%--------------------------------------------------------
For the RP1 model they can be written as
\begin{eqnarray}
\label{equation:A:RP1}
\nuU=\Omega_\K\Delta\,,\quad
\nuL=\left(1-\Gamma\right)\Omega_\K\,,
\end{eqnarray}
%--------------------------------------------------------
and for the RP2 model as
\begin{eqnarray}
\label{equation:A:RP2}
\nuU=\left(2-\Delta\right)\Omega_\K\,,\quad
\nuL=\left(1-\Gamma\right)\Omega_\K\,.
\end{eqnarray}
%--------------------------------------------------------

%----------------------------------------------------------
\subsection{Predictive Power of the RP Model}
%--------------------------------------------------------
\label{section:appendix:PP}

Let us assume a non-rotating star. The radial epicyclic frequency vanishes at ISCO, $x=6$, where the orbital frequency takes the value of
\begin{equation}
\nuK=\nu_{\mathrm{ISCO}}=\frac{c^3}{12\sqrt{6}\,G {M} \pi}
\end{equation}
and within the RP model it is
%--------------------------------------------------------
\begin{equation}
\label{equation:ISCO}
\nuU=\nuL=\nu_{\mathrm{ISCO}}\,.
\end{equation}
%--------------------------------------------------------
When a certain variation of the mass, $\delta\equiv \Delta M/M$, is assumed, the point in the frequency plane given by Equation (\ref{equation:ISCO}) changes its position. The corresponding square of the distance $ds^2$ (important for the fitting of data) reads
%--------------------------------------------------------------
\begin{equation}
ds_{\mathrm{ISCO}}^{\,2}=\nu_\mathrm{ISCO}\frac{\delta^2}{\left(1+\delta\right)^2}\,.
\end{equation}
%--------------------------------------------------------------
For any other specific orbit inside the accretion disk (e.g., $x=8$, where the radial epicyclic frequency takes its maximal value), the analogous change of the related data point position in the frequency plane is always smaller, $ds^{2}< ds_{\mathrm{ISCO}}^{\,2}$.

It is useful to utilize the fact that each specific orbit can be related to a certain frequency ratio $R$ higher than $R=1$  corresponding to ISCO (e.g., for $x=8$ it is $R=2$). Using the relation between $x$ and $R$ \citep[e.g.,][]{tor-etal:2008a}, one can find that
%--------------------------------------------------------------
\begin{equation}
\label{equation:A:PP1}
ds^2= ds_{\mathrm{ISCO}}^{\,2}\times\mathcal{P},
\end{equation}
%--------------------------------------------------------------
where
%--------------------------------------------------------
\begin{equation}
\label{equation:A:PP2}
%\mathcal{P}=\frac{\left(1-2R\right)^4 \left(\frac{3R^2}{2R-1}-1-2\sqrt{\frac{\left(R-1\right)^2}{2R-1}} \sqrt{\frac{R^2}{2R-1}}\right)}{2R^8}
\mathcal{P}=\frac{\left(R^2+1\right)\left(2R-1\right)^3}{2R^8}
\,.
\end{equation}
%--------------------------------------------------------
The quantity $\mathcal{P}=\mathcal{P}(R)$ reads $\mathcal{P}=1$ for $R=1$ and strongly decreases with increasing $R$. This naturally illustrates that the predictive power of the model is high only for orbits close to ISCO. For instance, for the maximum of the radial epicyclic frequency where $R=2$, it is roughly $\mathcal{P}=0.25$.

We note that in this subsection we neglected the influence of the NS spin for simplicity. Calculating $P$ for a non-zero $j$ is less straightforward and does not bring any new interesting information.

%--------------------------------------------------------
\subsection{Generalized Degeneracy}
%--------------------------------------------------------
\label{section:appendix:HT}
As recalled in Section~\ref{section:introduction:Mj}, the frequency curves predicted by the  model (and other kHz QPO models) scale with the NS mass and angular momentum, but do not change their shape much when $j\lesssim0.5$. This was explored in detail assuming the Kerr spacetimes. The exterior of a rotating NS is in general well described by the HT spacetimes which are determined by the NS mass $M$, angular momentum $j$, and a quadrupole moment $q$ reflecting the NS oblateness. One can ask whether there can be a \qm{generalized degeneracy} related to all these three quantities similar to those related just to $M$ and $j$ in the Kerr spacetimes. We briefly attempt to resolve this issue using formulas for epicyclic frequencies in HT spacetimes derived by \cite{abr-etal:2003a}.

The orbital frequency at a marginally stable circular orbit increases with increasing angular momentum $j$  while it decreases with increasing quadrupole moment $q$. Thus, following Appendix~A.2 of Paper~I, we can expect that the eventual generalized degeneracy can, to  first order in $q$ and  second order in $j$, be expressed as
%--------------------------------------------------------
\begin{equation}
\label{equation:HT:general}
M\sim M_0\left(1+{k}_1\,j+{k}_2\,j^2-{k}_3\,q\right)\,.
\end{equation}
%--------------------------------------------------------
In the limit of $\tilde{q}=1$, where $\tilde{q}\equiv{q/j^2}$ is the so-called  \qm{Kerr parameter}, relation (\ref{equation:HT:general}) has to merge with the mass spin relation derived for the Kerr spacetimes. This relation is represented by Equation (\ref{equation:general}) which, assuming whole frequency curves, reads
%--------------------------------------------------------
\begin{equation}
\label{equation:kerr07}
M\sim M_0\left[1+0.7\left(j+j^2\right)\right]\,.
\end{equation}
%--------------------------------------------------------
Therefore we choose ${k}_1=0.7$ and ${k}_2={k}_1+{k}_3$. Then only ${k}_3$ remains as a \qm{tunable} parameter.

We searched for a value of ${k}_3$ providing the eventual generalized degeneracy. For a particular choice of ${k}_3=0.32$,
%---------------------------------
\begin{equation}
\label{equation:HT:32}
M=M_0 \left(1+0.7j+1.02j^2-0.32 q\right)\,,
\end{equation}
%---------------------------------
we found results in full analogy to those that we had previously obtained for the Kerr spacetimes. This finding is illustrated in Figure~\ref{figure:EOSLTP}(a). The figure is plotted for  
$j\in(0,~0.3)$ and $\tilde{q}\in(1,~8)$. Clearly, for any curve drawn for a particular combination of $M$, $j$, and $q$ there is a nearly identical curve drawn for the Schwarzschild spacetime given by Equation~(\ref{equation:HT:32}).  Thus, consideration of NS oblateness cannot improve the poor quality of fits of models to the data within the limits of $j$ and $q$ assumed for the figure. These limits correspond to almost any NS modeled using the usual EoS for the mass $M>1.4M_\sun$ and spin frequencies up to 600Hz \cite[][]{lat-pra:2001,lat-pra:2007}.%\footnote{For completeness, we attempted to fit the data of 4U~1636-53 with the curves calculated in HT spacetime within the considered range of parameters and have not found any significant improvement in comparison to the fits in Kerr-spacetimes \issues{even when the upper limit for $j$ was increased up to $j=0.5$ (?)}.}

Considering the above facts, we can summarize the findings as follows: the results on $M-j$ relations obtained for the Kerr spacetimes have rather general validity and NS oblateness could only cause some correction to the slope of a particular $M-j$ relation. The only exceptions exceeding the framework of the  work presented are represented by the cases of $j\gg0.3$, $M<1.4M_\sun$, or some unusual NS models that have to be treated in detail assuming concrete EoS.

%----------------------------------------------------------
\section{Removing Degeneracy in the Case of the RP Model and 4U~1636--53}
%--------------------------------------------------------
\label{section:appendix:RP}
%\subsection{Spin Frequency Measurements}
%----------------------------------------------------------
\label{section:eos}
%----------------------------------------------------------
\emph{For the atoll source 4U~1636--53 there is good evidence on the NS spin frequency based on X-ray burst measurements. Depending on the (two- or one-) hot-spot model consideration, the spin frequency $\nu_\S$ reads either $\nu_\S\sim290$\,Hz or $\nu_\S\sim580$\,Hz \citep{str-mar:2002}. Thus, one can, in principle, infer the angular momentum $j$ and remove the $M-j$ degeneracies related to the individual twin peak QPO models.} 

In Figure~\ref{figure:EOSLTP} we illustrate the potential of  such an approach requiring a complex usage of various versions of a detailed ultra-dense matter description. The figure is made for the non-geodesic version of the RP model based on Equation~(\ref{equation:modification}) with $\beta\neq0$. It includes a $\chi^2$ map resulting from the fitting of 4U~1636--53 data with the model together with the $M-j$ relations inferred from the equalities $\nu_\S=290$\,Hz or $\nu_\S=580$\,Hz. These $M-j$ relations that depend on  ultra-dense matter properties were calculated using the approach of \citet{har:1967}, \citet{har-tho:1968}, \citet{cha-mil:1974}, \citet{mil:1977}, and \cite{urb-etal:2010:aca}. They assume the same set of several EoS as we used in Paper~I, namely
%\begin{itemize}
%\item
SLy~4~\citep{rik-etal:2003}, APR~\citep{akm-etal:1998}, AU-WFF1, UU-WFF2, and WS-WFF3~\citep{wir-etal:1988,ste-fri:1995}.

Comparing the $\chi^2$ map to the $M-j$ relations based on our choice of EoS we can conclude that the parameters of the NS implied by the model must be either $j\sim0.11$ and $M\sim1.9M_\sun$, or $j\sim0.22$ and {$M\sim2M_\sun$}. In panel (b) of Figure~\ref{figure:EOSLTP} we can check that in both cases the quality of fit to twin peak QPO data is acceptable (the best fits were obtained for the value of $\beta\sim0.15-0.2$).

\subsection{Adding Low-frequency QPOs}
%----------------------------------------------------------
\label{section:appendix:RP:low}
%----------------------------------------------------------

The RP model associates the observed low-frequency QPOs to the Lense--Thirring precession that occurs at the same radii as the periastron precession crucial for the high-frequency part of the model. It is then expected that their frequencies $\nu_\ell$ equal  the  Lense--Thirring precession frequency,
\begin{equation}
\nu_\ell=\nu_\mathrm{LT}.
\end{equation}
Naturally, the value of $\nu_\mathrm{LT}$ depends more strongly on the angular momentum $j$ than on the concrete radius $r$, since it vanishes for $j\rightarrow0$ at any radius. Thus, within the framework of the RP model, it represents a sensitive spin indicator \citep[][]{ste-vie:1998a,ste-vie:1998b,mor-ste:1999}. Although in this paper we focus on the high-frequency QPOs, it is interesting to mention this consideration, especially because of the relation to the above-mentioned implications of briefly X-ray burst measurements.  

There are several published observational works on QPOs in atoll sources including data points in the three-dimensional (3D) frequency space \mbox{$\mathcal{S}=\left\{\nu_\ell,~\nuL,~\nuU\right\}$}. For instance, \cite{jon-etal:2005} reported clear measurements of low-frequency QPOs in 4U~1636--53 as well as their relation to the high-frequency part of PDS. %The measured frequencies were approximately around $\nu_{\ell}=42$Hz for PDS with $\nuL=(700\div800)$Hz and $\nuU=1000\div1100$Hz or $\nu_{\ell}=43.5$Hz for PDS with $\nuL=(800\div850)$Hz and $\nuU=1100\div1150$Hz.
For the PDS related to the middle part of the frequency correlation,
%---------------
\begin{equation}
[\nuL,~\nuU]=[700-800\mathrm{Hz},~1000-1100\mathrm{Hz}],
\end{equation}
%---------------
the frequencies $\nu_\ell$ were approximately around 
%---------------
\begin{equation}
\nu_{\ell}\doteq42\mathrm{Hz}. 
\end{equation}
%---------------
For the PDS related to the upper part of the frequency correlation,
%---------------
\begin{equation}
[\nuL,~\nuU]=[800-850\mathrm{Hz},~1100-1150\mathrm{Hz}],
\end{equation} 
%---------------
the frequencies $\nu_\ell$ were around 
%---------------
\begin{equation}
\nu_{\ell}\doteq43.5\mathrm{Hz}.
\end{equation}  
%---------------

\noindent
Assuming these frequency intervals we can apply the equalities
\begin{equation}
\label{equation:fullRP}
\nuU=\nuK,~\nuL=\nu_\mathrm{RP}=\nuK-\nur\quad\mathrm{and}\quad\nu_\ell=\nu_\mathrm{LT}=\nuK-\nuv.
\end{equation}
For the application we consider Equation~(\ref{equation:modification}) with $\beta=0.17$ which provides acceptable fits to the twin peak QPOs. The spin $j$ is then  fixed just by the ratio between the observed frequencies (\ref{equation:fullRP}). Consequently we find that $j$ must be about $j=0.285-0.3$. Moreover, when using the measured frequency values, the relations~(\ref{equation:fullRP}) determine both $M$ and $j$ just for a single point in the 3D frequency space $\mathcal{S}$. Using this fact and the values of \cite{jon-etal:2005} we find that $M=(2.0-2.2)M_\sun$ for $j=0.285-0.3$.

The resulting values of $M$ and $j$ are marked in Figure~\ref{figure:EOSLTP} by the green box. Note, however, that the consideration needs to be further expanded for a larger set of data and some $\chi^2$ mapping in the 3D frequency space $\mathcal{S}$ should be done. This can be somewhat complicated by the fact that low-frequency QPOs are, in general, broader than the kHz features. In addition, the quadrupole momentum influence on $\nu_\mathrm{LT}$ could be overestimated due to the Kerr geometry approximation considered here. Nevertheless, assuming all these uncertainties we can still expect from the above numbers that a further detailed consideration should confirm the value of $j$ roughly inside the interval
\begin{equation}
%j_{\mathrm{LT}}=0.3(+0.005,-0.02).
j_{\mathrm{LT}}=0.3 \pm 0.05.
\end{equation}

{Figure~\ref{figure:EOSLTP} finally integrates both the implications of X-ray burst measurements and the Lense--Thirring precession model for low-frequency QPOs. We can see that an EoS relatively distant from those which we consider here could be needed in order to match both phenomena and fix the NS spin. This challenging issue clearly requires  further future work joining data analysis in the field of 3D frequency space and modeling the detailed influence of the NS EoS.}

%--------------------------------------------------------
%--------------------------------------------------------
%											REFERENCES                      
%--------------------------------------------------------
%--------------------------------------------------------

%---------------------------------------- 

%----------------------------------------
\end{document}